\documentclass[fleqn,usedcolumn,usenatbib]{mnras}

\usepackage{newtxtext}

\usepackage{longtable}

\usepackage[T1]{fontenc}
\usepackage{ae,aecompl}

\usepackage[british]{babel}             
\usepackage{newtxtext}                  
\usepackage[slantedGreek]{newtxmath}    
\let\la=\lesssim 
\usepackage{graphicx}                   

\usepackage[T1]{fontenc}
\usepackage{aecompl}
\usepackage{graphicx}	
\usepackage{amsmath}	
\usepackage{amssymb}	
\usepackage{pdflscape}  
\usepackage{multirow}
\usepackage{float}
\usepackage[dvipsnames]{xcolor}

\PassOptionsToPackage{fleqn}{amsmath}
\PassOptionsToPackage{british}{babel}

\title[Binary BHLA]{Bondi-Hoyle-Lyttleton accretion by binary stars}

\author[T.A.F.~Comerford et al.]{
T.A.F. Comerford$^{1}$\thanks{E-mail: tafc2@cam.ac.uk},
R.G. Izzard$^{2,1}$,
R.A. Booth$^{1}$ and
G. Rosotti$^{3,1}$
\\
$^{1}$Institute of Astronomy, Madingley Road, Cambridge, CB3 0HA, United Kingdom.\\
$^{2}$Astrophyhsics Research Group, Faculty of Engineering and Physics, University of Surrey, Guildford, GU2 7XH, United Kingdom.\\
$^{3}$Leiden Observatory, Leiden University, PO Box 9513, 2300 RA Leiden, The Netherlands.\\
}

\date{Accepted XXX. Received YYY; in original form ZZZ}

\pubyear{2019}

\begin{document}
\label{firstpage}
\pagerange{\pageref{firstpage}--\pageref{lastpage}}
\maketitle

\begin{abstract}
Binary stars often move through an ambient medium from which they accrete material and angular momentum, as in triple-star systems, star-forming clouds, young globular clusters and in the centres of galaxies.
A binary form of Bondi-Hoyle-Lyttleton accretion results whereby the accretion rate depends on the binary properties: the stellar masses and separation, and the relative wind speed.
We present the results of simulations performed with the hydrodynamic code \textsc{gandalf}, to determine the mass accretion rates over a range of binary separations, inclinations and mass ratios.
When the binary separation is short, the binary system accretes like a single star, while accretion onto stars in wide binaries is barely affected by their companion. 
We investigate intermediate-separation systems in some detail,
finding that as the binary separation is increased, accretion rates smoothly decrease from the rate equal to that of a single star to the rate expected from two isolated stars.
The form of this decrease depends on the relative centre-of-mass velocity of the binary and the gas, with faster-moving binaries showing a shallower decrease.
Accretion rates vary little with orbital inclination, except when the orbit is side-on and the stars pass through each others' wakes.
The specific angular momentum accretion rate also depends on the inclination but is never sufficient to prevent the binary orbit from contracting.
Our results may be applied to accretion onto protostars, pollution of stars in globular and nuclear clusters, and wind mass-transfer in multiple stellar systems.
\end{abstract}




\section{Introduction}

Bondi--Hoyle--Lyttleton accretion (BHLA) describes how a massive object moving relative to a uniform fluid gravitationally focusses and accretes material.
Based on the original ballistic model of \cite{HoyleLyttleton}, the BHLA rate accounts for gas pressure and shock formation, and smoothly interpolates between stationary and supersonic accretors \citep{BondiHoyle44,Bondi}. For a detailed review of BHLA see \cite{Edgar04}.
We The BHLA rate is applied to a variety of situations, ranging in scale from wind mass transfer in binary stars \citep{Boffin88} and the accretion of gas by stars in stellar clusters \citep{thoul02}, to accretion by entire stellar clusters \citep{LinMurray2007} and galaxies \citep{Sakelliou00}. In many such cases, the accretion rates predicted by theory closely match the observed consequences.

It is possible that the accretor in any of these contexts is a gravitationally bound binary, rather than single star system.
However, until recently, there has been little investigation of BHLA onto binary star accretors.
This is a problem for accurate modelling of mass transfer in triple star systems, in which a wind from a giant star encounters a binary in orbit around the giant.
In particularly close binaries, this could also pose issues if a similar triple system ever undergoes common envelope evolution in which all three stars orbit within a single envelope.
Modifications to accretion rates caused by binary accretors could also serve to alter predictions of stellar pollution in much the same way as wind accretion produces stars like barium and extrinsic carbon stars.

The Galactic halo object CS~22964-161 is a double-lined spectroscopic binary in which both stars are carbon and barium rich dwarfs \citep{Thompson_2008}. The system is likely a triple containing the relatively binary we see now, with a more distant white dwarf companion that transferred  carbon and barium to the binary in its wind when it ascended the asymptotic giant branch  \citep[AGB;][]{2016A&A...587A..50A}. 

\cite{Soker04} seeks to explain the shapes of irregular planetary nebulae as a result of BHLA onto a close binary in a triple system with an AGB star.
In this situation, they find that it is possible for the stars in the binary to accrete sufficient angular momentum to launch jets which shape the surrounding AGB wind, however only limited consideration is given to the mass accretion, its effect on the binary and its subsequent evolution.

We wish to explore the effects of binary BHLA in situations not studied in that work by relaxing the assumption that the accretion only occurs within a narrow column, and that the binary separation is much smaller than the Bondi-Hoyle radius,
\begin{equation}
    r_{\mathrm{a}} = \frac{2GM}{v^2}\,,
\end{equation}
where $M$ is the combined mass of the binary, and $v$ is its centre-of-mass velocity relative to the gas. For comparison, it is also convenient to define the Bondi radius,

\begin{equation}
    r_{\mathrm{B}} = \frac{G M}{c_s^2}\,,
    \label{eq:bondi-radius}
\end{equation}
where $c_s$ is the sound speed in the ambient gas.

\cite{LinMurray2007} consider the case of gas accretion onto a stellar cluster, modelled by a relatively shallow Plummer potential.
They find that under the condition that the cluster's internal velocity dispersion is slower than both its motion and the sound speed of ambient gas, bulk accretion by the cluster far outweighs the combined accretion rate of its stars. Most of this gas is retained inside the cluster and is not, at least initially, accreted onto stars.

Dimensional arguments show that accretion that is dominated by gravity has a rate, $\dot{M} \propto M^2$, where $M$ is the mass of the accreting object.
If a binary star has twice the mass of two otherwise identical single stars, this naively yields an extra factor of four in accretion rate.
The accreted mass is then split between two stars and each accretes at twice the rate compared to if it were isolated.
Naturally, this reasoning cannot hold in all binary systems, as some will have separations wider than the characteristic length scale of BHLA flow, but it is reasonable to assume that the accretion enhancement factor approaches four as the binary separation becomes arbitrarily short.

Because this problem is analytically intractable, we have carried out a suite of smoothed particle hydrodynamic (SPH) simulations using the code GANDALF \citep{GANDALF}.
Our simulations probe a range of binary separations, velocities and inclinations, with the aim of determining a numerical prefactor to the analytic accretion rate which depends on the binary properties.

\section{Analytic Prescriptions}
\label{sec:analytic}

\begin{figure}
    \centering
    \includegraphics[width=0.75\columnwidth]{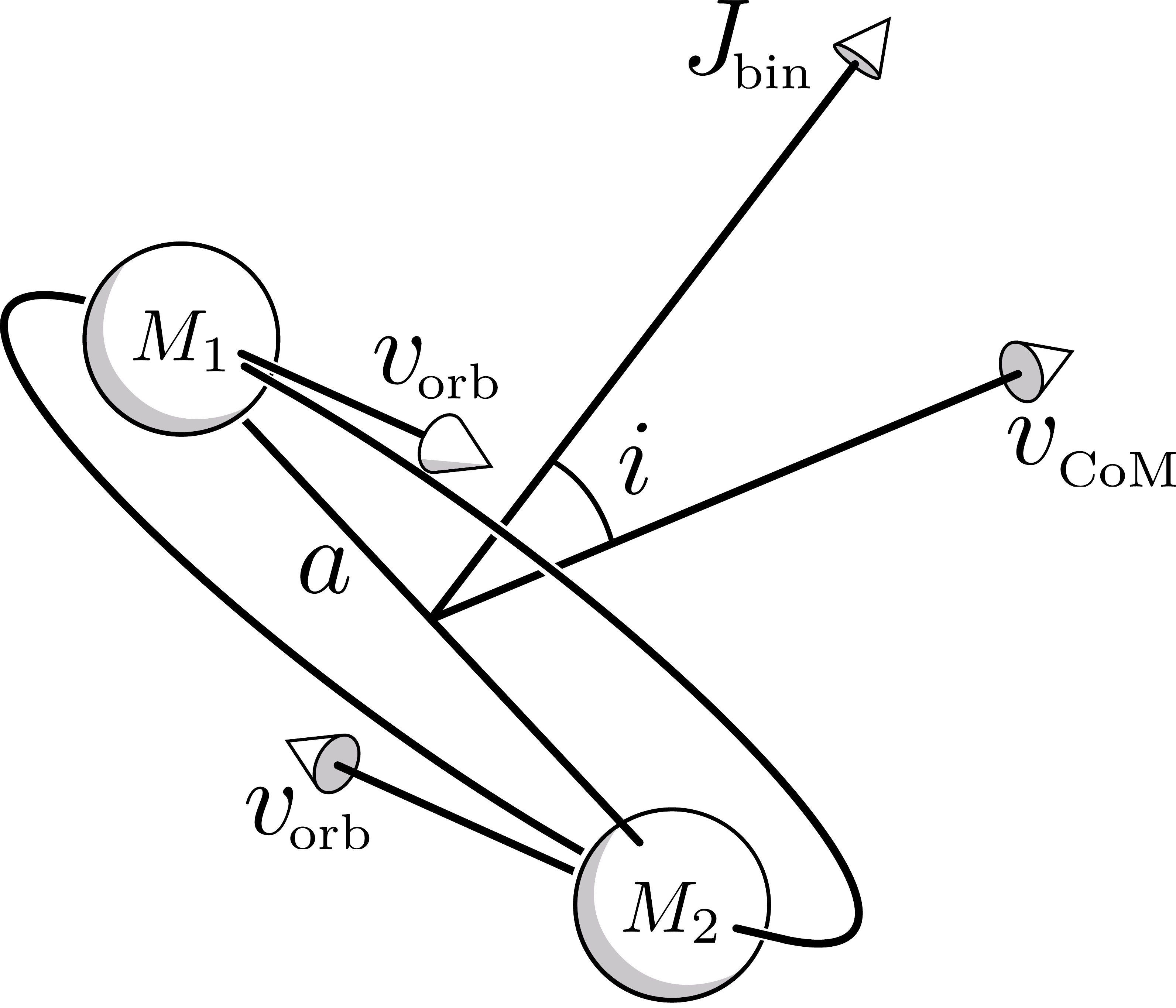}
    \caption{A schematic of a typical binary system considered in this paper. $v_{\mathrm{orb}}$ represents the binary's orbital velocity, while $v_{\mathrm{CoM}}$ is the centre-of-mass velocity relative to the surrounding gas. Note that this figure shows an equal-mass binary, while we also consider binaries with unequal masses (see \S \ref{section:unequal-mass}).}
    \label{fig:system}
\end{figure}

The original description of \cite{HoyleLyttleton} considers a star, modelled as a point mass, moving relative to a uniform ideal fluid at a constant, supersonic velocity.
The gravity of the star focuses passing material into a one-dimensional shock in its wake.
As gas decelerates into the shock, it loses speed in the direction perpendicular to the wake. This deceleration causes some material to become gravitationally bound to the star, i.e.~it will eventually be accreted.
The initial impact parameter at which material will become marginally bound is the Hoyle-Lyttleton radius, $r_\mathrm{a}$.
By considering the rate at which incoming gas with impact parameter less than $r_\mathrm{a}$ approaches the star, the accretion rate is,
\begin{equation}
    \dot{M}_\mathrm{HL} = \pi r_\mathrm{a}^2 \rho v = 4\pi\rho \frac{G^2 M^2}{v^3}\,,
    \label{eq:hl-rate}
\end{equation}
termed the Hoyle-Lyttleton accretion rate. Here, $\rho$ is the ambient gas density, $M$ is the mass of the star, $v$ is the relative velocity of the star and gas, and $G$ is the gravitational constant.

This description suffers from several drawbacks, the most serious being that the accretion column has multivalued velocity: material must simultaneously flow towards and away from the star. It also treats the fluid as collisionless, except in the shock.

A more realistic model was proposed in \cite{BondiHoyle44} which introduces a finite shock width. A three-dimensional accretion column avoids the problem of multivalued velocity, although it is still assumed that the fluid moves ballistically outside the shock.
A convenient property of this model is that the accretion rate at low relative velocities approaches the form of the Bondi rate,
\begin{equation}
    \dot{M}_\mathrm{B} \approx 2 \pi \rho \frac{G^2 M^2}{c_\mathrm{s}^3}\,.
    \label{eq:b-rate}
\end{equation}
This results in an accretion rate which interpolates between the Bondi rate and the supersonic Hoyle-Lyttleton rate,
\begin{equation}
    \dot{M}_\mathrm{BHL} = 4 \,\times\, \pi \rho \frac{G^2 M^2}{\left(v^2 + c_\mathrm{s}^2\right)^{3/2}}\,,
    \label{eq:bhl-rate}
\end{equation}
known as the Bondi-Hoyle-Littleton (BHL) rate. While \ref{eq:b-rate} implies that the numerical constant should be 2, it is not particularly well constrained by the treatment in \cite{Bondi}.
So, following the numerical simulations in \cite{Shima85}, we set it to 4 to match the hypersonic, i.e.~large $v$, Hoyle-Lyttleton rate (Eq.~\ref{eq:hl-rate}).

In systems with binary accretors we expect three regimes of accretion depending on the ratio of binary orbital separation, $a$, and the typical accretion radius, $r_\mathrm{a}$.
If $a \gg r_\mathrm{a}$, the system behaves similarly to two single stars and accretion flows should be of the standard BHLA form onto each star.
In this case, one would expect the accretion rate to be,
\begin{equation}
    \dot{M} = \dot{M}_1 + \dot{M}_2 = \left[\frac{q^2}{(1+q)^2} + \frac{1}{(1+q)^2} \right] \dot{M}_{\mathrm{BHL}}~,
    \label{eq:separate-rate}
\end{equation}
where $\dot{M}_{\mathrm{BHL}}$ is the BHL accretion rate for a star with a mass equal to $M_1 + M_2$ (which we term the `single-star rate').

When $a \ll r_\mathrm{a}$, the accretion flow should be similar to a point mass containing the total mass of the binary at the BHL rate, with deviations only occurring close the the stars.
In this case, one might expect that accreted mass is divided between the stars in the same ratio as Bondi accretion, with the total accretion rate set by the point-mass BHL rate ($\dot{M}$ in Eq.~\ref{eq:bhl-rate}). Alternatively, if the accreted material has sufficient angular momentum, circumstellar or circumbinary discs may form (a general investigation of BHLA onto a disc appears in \citealt{moeckel09}).

In the intermediate case, where $a \sim r_\mathrm{a}$, it is unclear which regime should dominate, as the two characteristic velocity scales, $c_{\mathrm{s}}$ and the binary's orbital velocity,
\begin{equation}
    v_{\mathrm{orb}} = \sqrt{\frac{GM}{a}},
\end{equation}
are approximately equal; that is to say,
\begin{equation}
    \frac{r_{\mathrm{a}}}{a} \approx \frac{v_{\mathrm{orb}}^2}{\sqrt{v^2 + c_s^2}}~.
    \label{eq:distance-velocity-scaling}
\end{equation}
As a result, we have chosen to perform the majority of our hydrodynamic simulations on systems in the range of orbital separation where the ratios in Eq.~\ref{eq:distance-velocity-scaling} are of order unity.

To first order, we expect the intermediate-case accretion rate to exceed that of wide binaries because of the proximity, and added gravitational pull, of the companion.
However, closer binaries have higher orbital velocities, which may disrupt the accretion flow and subsequently reduce the accretion rate.
This could be counteracted if the binary retains gas that is not bound to either star.
In this case, the gas density near the individual stars will be increased, potentially offsetting the decrease due to orbital velocity.

The stability of BHLA has is the topic of some debate, with conclusions depending on dimensionality and assumed symmetries, as well as physical quantities such as Mach number, $\mathcal{M}$, and accretor size \citep{Foglizzo05}.
A variety of instability mechanisms and types are have been proposed, generally sharing the property that at low Mach numbers the flow is stable, only becoming unstable for $\mathcal{M} \gtrsim 10$ or accretors with sizes much less than the accretion radius.
Because none of our simulations fall in this range, we do not observe nor further address instability of the flow in this paper.

\section{Hydrodynamics}

We simulate hydrodynamics with the GANDALF $\nabla h$ smoothed particle hydrodynamics (SPH) code \citep{GANDALF} using an M4 cubic spline kernel \citep{M4kernel} and a leapfrog kick-drift-kick integration scheme.
To accurately simulate energy dissipation in shocks we solve the energy equation for an ideal gas, using an adiabatic index of 5/3, and without artificial conductivity.
Because we expect strong shocks, we use the \cite{artificial-viscosity} alpha-viscosity, with a viscosity switch as described in \cite{viscosity-switch}. This prevents particles jumping across a shock in a single simulation timestep and ensures that the shocks are accurately simulated.
We do not include the effect of self-gravity of the gas, nor the subsequent dynamical friction on the stars.
This reduces the computational requirements, as well as keeping our results independent of the gas density, and is in accord with the standard BHLA prescriptions in which the density of the gas is assumed to be negligible.

\cite{barai11} show that spherical Bondi accretion can be accurately modelled using the SPH code \textsc{gadget-3}, and at similar resolution to our simulations.
They note two unphysical effects that occur as a result of the numerical method and its finite resolution.
The first is excess heating of the gas near the accretor as a result of artificial viscosity. 
This effect should be diminished in our simulations because of the use of a viscosity switch, which reduces the artificial viscosity in regions away from shocks.
The switch allows the artificial viscosity to decay towards a minimum specified value in regions where the flow's velocity has positive or zero divergence.
For a radial inflow, if the velocity is locally described by a power law $v \propto r^{-\alpha}$, the condition for negative divergence is $\alpha > 2$.
Our simulations show $\alpha$ is typically close to 1 in the vicinity of the accretor (Appendix~\ref{app:resolution-study}), so the viscosity should still be its minimum value here.
That is not to say that there will be no heating from viscosity, but that the amount should be less than observed in \citet{barai11}.

The second unphysical effect produced in their simulations is a flow of material across the simulation boundary which occurs as a result of their initial setup and boundary conditions.
We avoid this problem by initially filling the entire domain of the simulation with uniform fluid and using periodic boundary conditions.

Our simulations begin with a cuboid of fluid at uniform density and temperature, and SPH particles in a cubic lattice.
The particle separation is linked to the smoothing length by the parameter $\eta$, where,
\begin{equation}
    \label{eqn:smoothing-length}
    h_i = \eta \left(\frac{m_i}{\rho_i}\right)^{1/3},
\end{equation}
and $h_i$, $m_i$, and $\rho_i$ are the smoothing length, mass, and density of particle $i$.
In all our simulations, we keep $\eta$ at its default value of 1.2, meaning that, initially, the smoothing length of the particles is 1.2 times their separation.
Most simulations use approximately 2, 4 or 8 million particles with a particle mass of $3 \times 10^{-8}$ in simulation units as defined in \S~\ref{sec:units}.
A full list of simulation runs is given in Table~\ref{tab:parameters-table}.
The stars are treated as sink particles with a radius equal to the smoothing length of the surrounding particles.
We do not model finite stellar size, magnetic fields, stellar winds or stellar evolution.

As GANDALF also supports a meshless finite volume (MFV) scheme based on \cite{MFV}, we compare the two different methods.
However, because the MFV simulations required approximately 10 times more computation time than SPH, we perform only a small number for comparison to our SPH results.
\cite{Hopkins15} show that in the case of a linear travelling sound wave, the numerical error of the MFV simulation scales similarly to SPH, and is always less than the SPH simulation with the same particle spacing due to its lower numerical viscosity. We perform both sets of simulations with the same particle spacing, since this means that the SPH and MFV simulations are similarly resolved.

\subsection{Parameter space}

Our simulations span a range of binary separations, inclinations, mass ratios and velocities. We do not treat eccentric binaries in this work because this would require two additional parameters, the eccentricity and longitude of periapsis.
Two main effects limit the ranges of parameters used in the simulations. The first is that if the binary separation is too wide, the size of the simulation box needed to prevent interaction with the boundary becomes impractically large.
The nature of our simulation setup also precludes simulations of very high-velocity binaries, because these will generally leave the simulation domain before reaching a steady state.

The parameter space is also limited by the finite resolution of our simulations. As shown in \cite{Ruffert94}, when the size of the accretor is large, the bow shock becomes attached to the accretor, producing accretion rates higher than in better-resolved simulations.
Overall, this effect results in the constraint that the accretor size (and so, the smoothing length) must be smaller than $r_{\mathrm{a}}$ by a factor that depends on the accretor's velocity. Appendix~\ref{app:resolution-study} shows the effects of varying resolution on our accretion rates.
This imposes an upper limit on the centre-of-mass velocity, and lower limits on the binary separation and mass ratio.
There are no such constraints on the inclination, so we perform simulations with inclinations of 0 (face on), 45, and $90^{\circ}$ (side on), to capture the range of behaviour without redundant simulations.
In addition to the binary simulations, we also carry out a series of single-star runs.
These serve as a comparison to the analytic rates (Eqs.~\ref{eq:hl-rate}--\ref{eq:bhl-rate}) and results in other works, and provide a baseline to compare the effect of binary accretors.
We do not vary the total mass of the system in our simulations, so all binary accretors have a combined mass equal to that of the single stars.

Because stars accrete SPH particles with finite mass, our measured accretion rates are also subject to shot noise.
Assuming a Poisson distribution, the expected relative error in accretion rate is,
\begin{equation}
    \sigma_{\mathrm{shot}} = \frac{1}{\sqrt{N}} = \sqrt{\frac{\left(v^2 + c_s^2\right)^{3/2} h^3}{3 G^2 M^2 \delta t}}~,
\end{equation}
where $N$ is the number of particles accreted during time $\delta t$ over which the accretion rate is measured.
For the accretion rate to not be significantly affected by shot noise, we impose the condition,
\begin{equation}
    h^3 \ll r_{\mathrm{a}}^2 \sqrt{v^2 + c_s^2}~\delta t~,
\end{equation}
which, when $\delta t$ is similar to the time taken for an accretor to cross its own BHL radius, reverts to the condition $h \ll r_{\mathrm{a}}$, which is similar to the condition that ensures that the accretion is well-resolved.
Since $\sigma_{\mathrm{shot}} \propto (\delta t)^{-1/2}$, shot noise affects the orbit-averaged accretion rates less than the derived instantaneous rates.

The dimensions of the parameter space are reflected in the numbering of the simulations in Figures~\ref{fig:snapshots} and \ref{fig:snapshot-rates}, and in Table~\ref{tab:parameters-table}.
The first digit refers to the mass ratio of the simulated binary, the second to its inclination, the third to its separation, and the fourth digit to the centre-of-mass velocity.
Note that the labelling of velocities is not unique, as more than ten different velocities were used in total.

\subsection{Processing}

From each of our simulations we extract the time-varying accretion rate of each accreting star.
To calculate the mean accretion rate we average over a suitable time interval.
In simulations with constant accretion rates (single stars and face-on systems), the time interval over which we average starts when the accretion rates reach their steady state, and ends at the end of the simulation, before the binary encounters the simulation boundary.
In simulations with time-varying accretion rates, the time interval is truncated at the end so that it is equal to an integral number of binary orbital periods.
This is done to avoid any offset in the average rate due to variation with orbital phase.
Typically, accretion rates reach a steady state after the accretor has travelled a distance equal to its BHL radius, $r_\mathrm{a}$, or the sound crossing time of the BHL radius, in the case of stationary accretors with $v=0$.
In face-on ($i=0^{\circ}$) systems, the accretion rate remains constant for the duration of the simulations, as one would expect from the symmetry of the system.
Inclined systems have accretion rates that vary with time. The most extreme cases are side-on ($i=90^{\circ}$) orbits for which the instantaneous accretion rate varies by up to a factor of two during an orbit.
Figure~\ref{fig:snapshot-rates} shows the accretion rates as a function of orbital phase for a variety of different binary configurations.

\subsection{Simulation units}
\label{sec:units}

Our simulations use dimensionless variables and our results are converted to physical quantities by multiplying by appropriate scale factors. A physical quantity, $x$, can be expressed as the product $X_0 x'$, where $X_0$ is a scale factor with the same units as $x$, while $x'$ is the dimensionless quantity used in the simulations.

We define our simulation units such that the simulation mass unit $M = \mathrm{M}_{\odot}$, and the isothermal sound speed is unity, $c_\mathrm{s} = \sqrt{kT/\mu} = 1$, where $\mu$ is the mass of an H$_2$ molecule. As a result, the scale factors depend on the temperature of the gas, $T$.

The resulting scale factors for velocity, length, time and accretion rate are,
\begin{subequations}
\begin{align}
V_0 &= 645 \left(\frac{T}{100\text{ K}}\right)^{1/2} \text{ m s}^{-1}, \\
R_0 &= 3.18 \times 10^{14} \left(\frac{T}{100\text{ K}}\right)^{-1} \text{ m}, \\
T_0 &= 4.93 \times 10^{11} \left(\frac{T}{100\text{ K}}\right)^{-3/2} \text{ s}, \\
\dot{M}_0 &= 5.05 \times 10^{19} \times \left(\frac{T}{100\text{ K}}\right)^{3/2} \text{ kg s}^{-1},
\end{align}
\end{subequations}
\noindent respectively. When $T = 100 \text{K}$, the scale factors for length, time and mass accretion rate are approximately $2020~\text{AU}$, $15650~\text{yr}$, and $8.03 \times 10^{-4}~\text{M}_{\odot} / \text{yr}$, respectively.

\section{Results}

In this section we present the results of our simulations by considering the initial parameter space one dimension at a time.
We start with a general overview of the phenomena observed in the simulations (\S \ref{sec:phenomenology}) and compare the SPH and MFV results (\S \ref{sec:comparison}).
We then analyse the accretion rates from single stars (\S \ref{sec:single-star}), then face-on equal-mass binaries (\S \ref{sec:face-on-twins}), inclined equal-mass binaries (\S \ref{sec:inclined-twins}), and finally the general problem with varying velocity, separation, inclination, and mass ratio (\S \ref{section:unequal-mass}).

\subsection{Phenomenology}
\label{sec:phenomenology}

\begin{figure*}
    \centering
    \includegraphics[width=\textwidth]{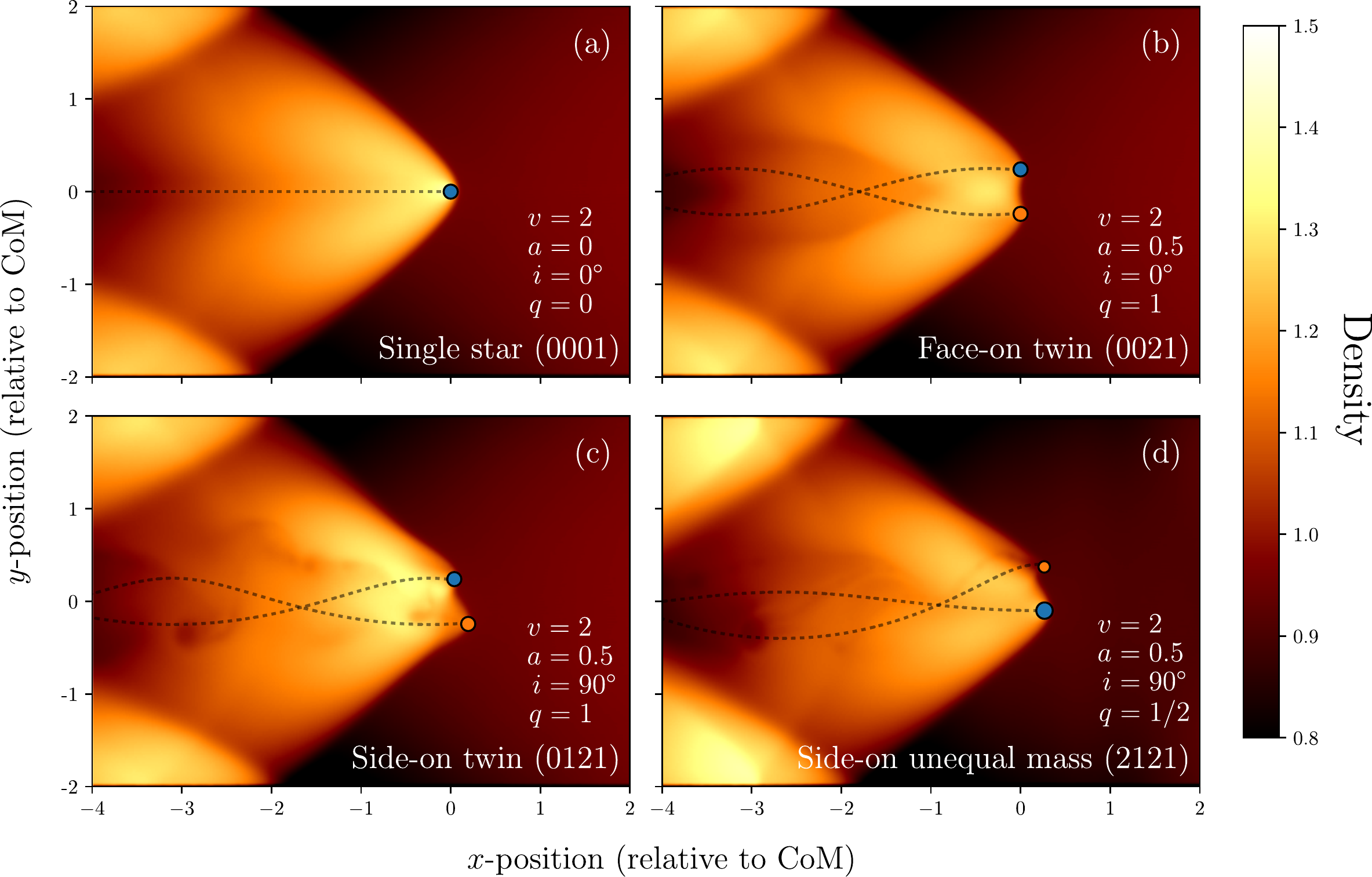}
    \caption{Sample snapshots from four of our simulation runs, labelled with the simulation parameters (also available in Table \ref{tab:parameters-table}, to which the simulation numbers refer). The accretors and their previous trajectories are marked with circles and dashed lines, respectively; the colours correspond to the line colours in Fig~\ref{fig:snapshot-rates}. The colour shows the gas density, integrated along the $z$-axis of the simulation domain and normalised such that the ambient density is $1$. The snapshots are taken at different times in each simulation, in order to best demonstrate the observed phenomena; see Fig~\ref{fig:snapshot-rates} for the orbital phase of each snapshot.}
    \label{fig:snapshots}
\end{figure*}

In each simulation, we observe the typical features expected of BHLA: a bow shock is formed ahead of the star(s), with a region of higher density gas behind (Figure \ref{fig:snapshots}).
In this region, fluid particle trajectories curve toward the accretor, such that the flow is almost radial near the accretor.
Because the simulations are carried out in a finite domain, the bow shock interacts with the simulation boundary downstream of the accretor(s). This is visible in Figure \ref{fig:snapshots} as the high-density regions at the top and bottom of each plot.

Among supersonic accretors, the interaction with the simulation boundary has no effect on the accretion rate; this was verified by performing identical simulations in wider boxes.
However, a slight change in accretion rate was observed in the subsonic accretors, even though simulation boxes are always at least twice as large as the characteristic accretion radius.
The varying accretion rate appears to be the result of a periodic change in density, varying with a period equal to the simulation's sound-crossing time.
The variation is most extreme in simulations with single accretors (i.e. standard Bondi accretion), where the instantaneous accretion rate varies by about 5 per cent.
We find that the variation is not present in the larger simulation boxes, although the mass resolution is subsequently lower, resulting in increased shot noise in the accretion rate.

\begin{figure}
    \centering
    \includegraphics[width=\columnwidth]{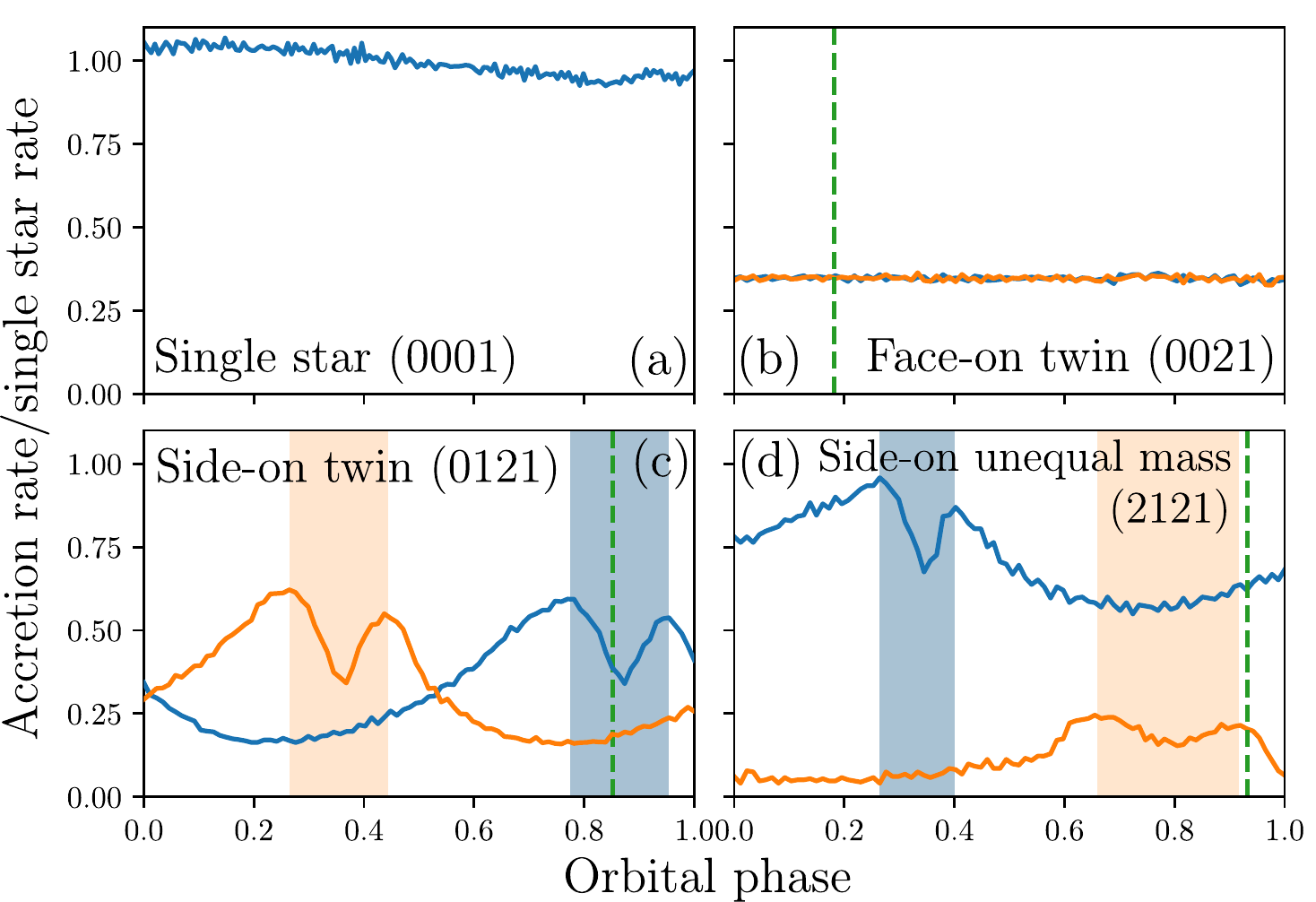}
    \caption{Accretion rates as a function of orbital phase in the four simulations shown in Fig \ref{fig:snapshots}. Accretion rates are normalised to the mean single star rate (subplot $a$), and the accretion rates for both stars are plotted individually where applicable. The vertical line shows the orbital phase at which the snapshots in Fig.~\ref{fig:snapshots} were taken. The shaded segments in plots (c) and (d) indicate the period when one star is passing through the wake of the other star. The colour of the shading indicates which star is currently passing through a wake, corresponding to the colours of the line in this figure and Figure~\ref{fig:comparison}, and the dots in Figure~\ref{fig:snapshots}.}
    \label{fig:snapshot-rates}
\end{figure}

Our higher-resolution simulations, in which accretion is better resolved, resulted in an accretion rate around 10 per cent higher than our original stationary simulation.
Because this factor does not appear to depend on binary separation, we apply it as a correction to the accretion rates of all stationary accretors.

Our SPH simulations also exhibit a wake of low density, high temperature gas.
Over time, the wake collapses into a series of beads, visually similar to the Plateau-Rayleigh instability \citep{papageorgiou95}.
The existence of the wake appears to have little effect on the accretion, except when one star encounters the other's wake in side-on systems.
In these cases, the instantaneous accretion rate drops by up to 50 per cent before quickly returning to normal (subplots $(c)$ and $(d)$ in Fig.~\ref{fig:snapshot-rates}).
However, because the duration of wake interaction is only up to 10 per cent of the orbital period, the time-averaged accretion rates are changed only by a few percent, which is of similar order to the shot noise in the measured accretion rates.

\subsection{Comparison between SPH and MFV}
\label{sec:comparison}

\begin{figure*}
    \centering
    \includegraphics[width=\textwidth]{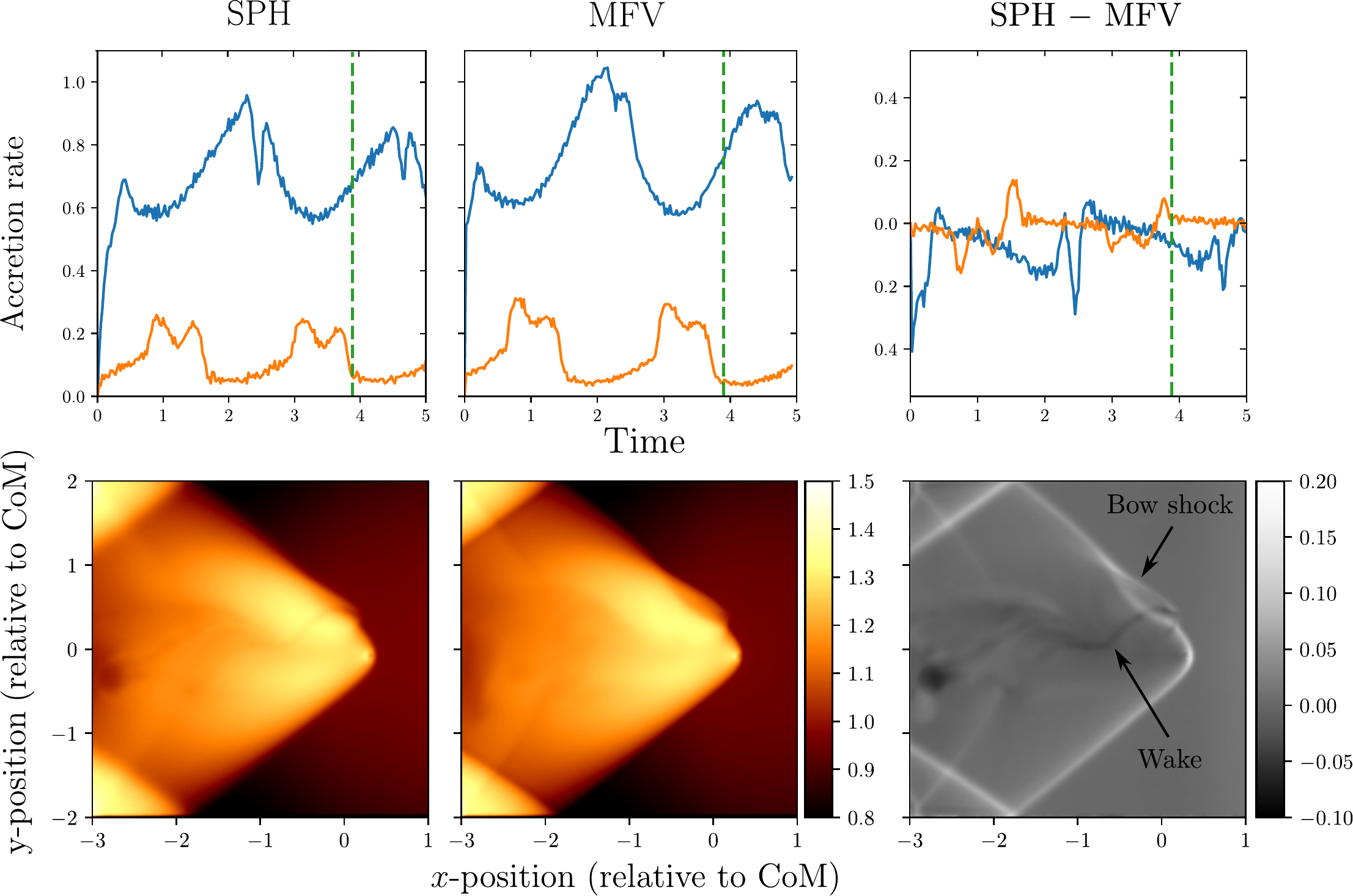}
    \caption{Comparison between our SPH and MFV simulations. SPH data are shown in the left-hand column, MFV in the centre, and the difference between SPH and MFV on the right. The top row shows accretion rates and the bottom row shows density snapshots. All units and symbols are as in Figures \ref{fig:snapshots} and \ref{fig:snapshot-rates}.}
    \label{fig:comparison}
\end{figure*}

We performed a small number of simulations using the MFV fluid scheme to determine whether our results are affected by SPH artefacts, such as surface tension \citep{springel10}.
Because of the more computationally intensive nature of the MFV simulations, we restricted this to the more complex simulations, namely those in which one star interacts with the other's wake.

The MFV simulations display very similar flow structure to the SPH, although there are some slight differences in the density structure between the two simulations (Figure \ref{fig:comparison}).
The post-shock region in the SPH simulations is approximately 10 per cent denser than in MFV, while the low-density wake observed in our SPH simulations is much less pronounced with MFV, and none of the beading occurs. 
We interpret the beading in the wake as being due to the surface tension between SPH particles with differing specific entropy, which artificially prevents mixing.
The low-density wake has little effect on the accretion rates, except when $i=90^{\circ}$ (side-on). In this case, the stars pass through the wake, creating dips in the accretion rate (\S\ref{sec:inclined-twins}).

\subsection{Single star accretion rates}
\label{sec:single-star}
\begin{figure}
    \centering
    \includegraphics[width=\columnwidth]{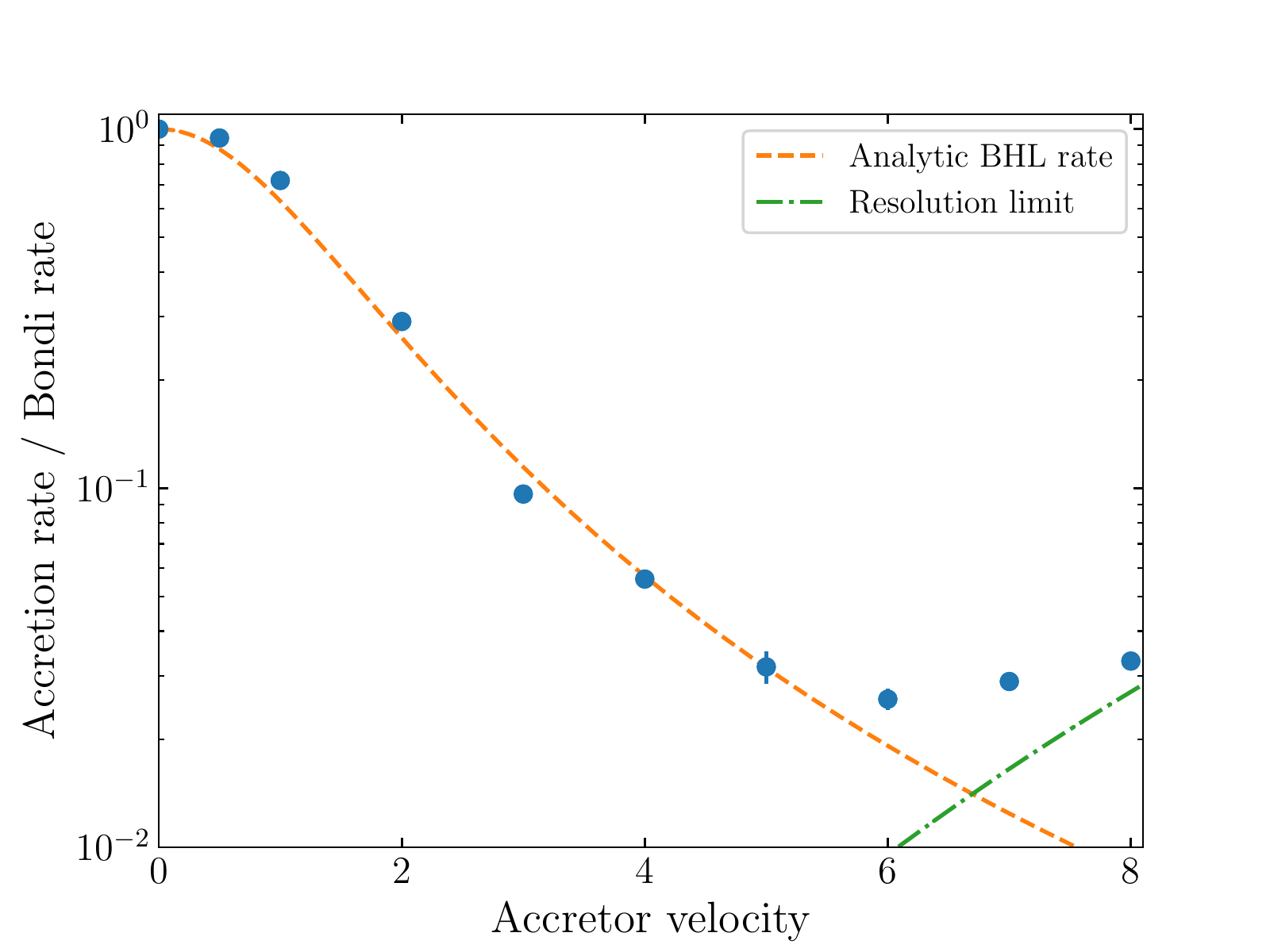}
    \caption{Accretion rates in our single stars versus velocity (in simulation units), normalised to the stationary Bondi accretion rate.
    The dashed orange line is the analytic BHLA rate (\ref{eq:bhl-rate}), and the green dot-dashed line is the expression for the unresolved accretion rate derived in \citet{Ruffert94}.
    The error bars show the standard deviation in our computed accretion rate, but are smaller than the points in most of our simulations.
    See Appendix~\ref{app:resolution-study} for a more detailed study on the impact of varying simulation resolution on our results.}
    \label{fig:single-star-rates}
\end{figure}

Our simplest set of simulations concern accretion onto a single sink particle as a test of the standard BHLA formalism. The relative velocity is ranged between $0$ and $8$ in simulation units, probing stationary ($v=0$), and supersonic ($v>1$) accretion, as well as the transition region between. The resulting accretion rates (Fig.~\ref{fig:single-star-rates}) follow the analytic BHLA rate (Eq.~\ref{eq:bhl-rate}) until $v = 6$, where the accretion becomes unresolved.
Faster than this, the simulation resolution is insufficient to resolve the bow shock, and the resulting accretion rate is simply proportional to the accretor velocity.
For this reason, we restrict our parameter space to $v\leq4$ in subsequent simulations.
From here on, we refer to these accretion rates as the `single star rates'.

\subsection{Face-on twins}
\label{sec:face-on-twins}
The second dimension of the parameter space is the binary semi-major axis, $a$, which we vary between $0$ and $1$, while the inclination is held at $0$ (accretion face-on to the orbit) and the mass ratio at $1$ (equal mass stars, each being half the mass of the stars in \S \ref{sec:single-star}).
We refer to binaries with these properties as `face-on twins'.
This is the simplest case of binary BHLA, as the accretion rate is constant with orbital phase, and is the same on to both stars.

\begin{figure}
    \centering
    \includegraphics[width=\columnwidth]{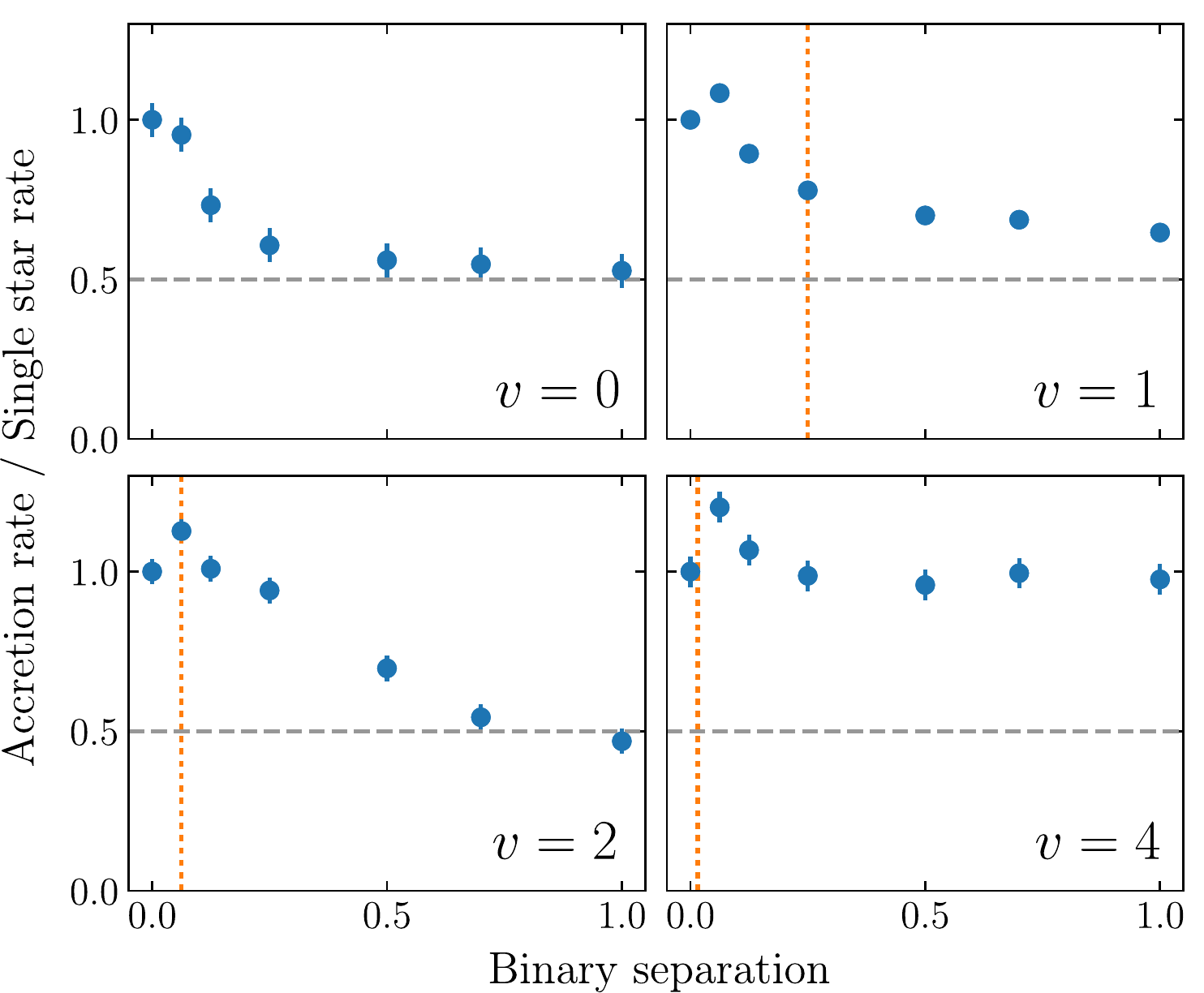}
    \caption{Accretion rates in all our face-on twin simulations at four different velocities. Each rate is normalised to the corresponding single-star ($a = 0$) rate. The horizontal dashed lines show an accretion rate ratio of $0.5$, which is the expected value at large separation. The vertical dotted lines show the binary separation at which the orbital velocity equals the centre-of-mass velocity. To the left, orbital velocity dominates the total velocity of each accretor; to the right, centre-of-mass velocity dominates.}
    \label{fig:face-on-twin-rates}
\end{figure}

The accretion rates are plotted in Figure \ref{fig:face-on-twin-rates}, separated by centre-of-mass velocity.
In most simulations, the accretion rate relative to the equivalent single-star rate tends to $1$ at small separations and to $1/2$ at wide separations, where the binary can be treated as a single accretor, and two non-interacting accretors, respectively.
However, this does not hold for binaries with $v=4$; these show a slight increase over the equivalent single-star rate for close binaries, and the ratio drops to 1 when $a / r_{\mathrm{a}} \geq 1/4$.
This appears to be the result of unresolved accretion, where, as $\dot{M} \propto v$, there increase accretion in close binaries owing to their high orbital velocities, which are non-negligible compared to total velocity of the binary.

The separation dependence of the accretion rate depends on the binary's velocity relative to the gas; in slower binaries, the accretion rate drops to half the single-star rate at shorter binary separations.
This can be explained by considering the total velocity of each accretor.
Because, in face-on binaries, the centre-of-mass velocity and orbital velocity of the individual accretors are perpendicular, they are added in quadrature to obtain the magnitude of the total velocity of each star relative to the gas.
As a result, at a given separation, the difference between single and binary accretors is most pronounced at slow centre-of-mass velocities, and the effect of duplicity is more pronounced than in fast accretors.
In Figure \ref{fig:face-on-twin-rates}, this is indicated by the dotted vertical lines. In the region to the left of the dotted line, the velocity of each individual accretor is dominated by the binary orbital velocity; to the right, the centre-of-mass velocity dominates.

There is also an increase in accretion rate observed in very close ($a \lesssim 0.1$) binaries, giving an accretion rate up to 20 per cent higher than in our single star simulations.
There are two possible explanations for this phenomenon. First, that this is a genuine physical phenomenon and that the higher rate could be justified by considering that the accretors are sweeping out a larger volume per unit time.
In effect, the accretors' motion effectively expands the accretion radius, $r_\mathrm{a}$, giving a higher mean accretion rate.
The second explanation, which we believe to be more likely, is that the increase results from unresolved accretion.
In less massive, faster accretors, the BHL rate decreases, but the limit at which accretion is unresolved increases.
The combination of these two effects enables the unresolved accretion to noticeably affect the accretion rate.
The magnitude of this effect is difficult to estimate, however, because the motions of the fluid and accretors are much more complex than the single-star case.

\subsection{Inclined twins}
\label{sec:inclined-twins}
\begin{figure}
    \centering
    \includegraphics[width=\columnwidth]{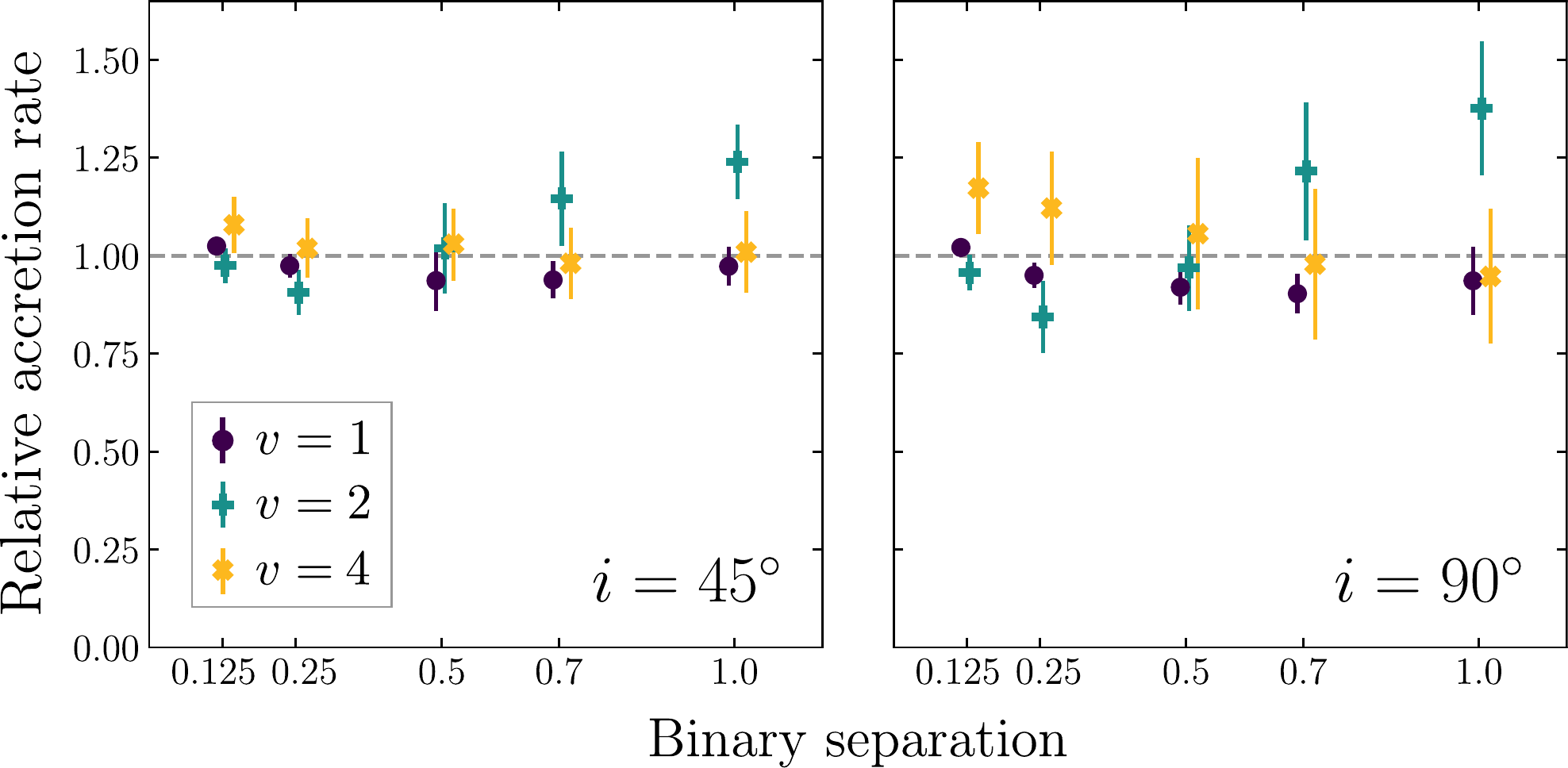}
    \caption{Accretion rates in all our inclined twin simulations: each rate is normalised to the face-on accretion rate. The left subplot includes all simulations with inclinations of $45^{\circ}$, and the right, simulations with $i=90^{\circ}$. The $x$-axis plots the binary separation, while the colour shows the centre-of-mass velocity in simulation units. As before, the bars show the standard deviation in accretion rate over one orbit.}
    \label{fig:inclined-twin-rates}
\end{figure}

When comparing our accretion rates at arbitrary inclinations to our face-on rates, the dependence of accretion rate on inclination is not particularly strong; the greatest deviation from the face-on case is 37 per cent.
While our mean accretion rates are similar at inclinations of $45^{\circ}$ and $90^{\circ}$, the generally larger bars in the right-hand plot of Figure~\ref{fig:inclined-twin-rates} indicate that the side-on accretion rates  vary more over an orbital period.
This is expected because stars in side-on ($i=90^{\circ}$) binaries encounter the other star's bow shock more often, and have a total velocity which varies more extremely than when face-on or at $45^{\circ}$.

From Figure~\ref{fig:inclined-twin-rates}, accretion rates are most enhanced in slow, but not stationary, wide binaries.
This is perhaps to be expected, because in an inclined binary stars spend considerable time moving through each other's high-density shocked region.
In slow accretors, this high-density region has a greater extent because of the large opening angle of the bow shock, while in wide binaries, the density enhancement due to proximity of the companion is weakest.

Compared to the face-on simulations with $v = 2$, there also appears to be an increase in accretion rate in both inclined binaries.
This is due to the spuriously low rates measured for simulations with $a = 0.7$ and $a = 1$ (Figure~\ref{fig:face-on-twin-rates}); when the absolute accretion rates are considered instead, these results are consistent with the rates at other binary separations.

\subsection{Varying mass ratio}
\label{section:unequal-mass}
The analytic prescriptions in \S\ref{sec:analytic} show that when the mass ratio is decreased, the total accretion rate should increase, tending toward the single-star rate as $q$ tends to zero.

The simulation results support this prediction; we find that the accretion rates monotonically increase with decreasing $q$, and that the results when $q=1/4$ are indistinguishable from the single-star rates.
Binaries with $q=1/2$ have accretion rates between 0 and 20 per cent higher than the equal-mass ($q=1$) rates; the rates in binaries with $q=1/4$ are enhanced between 0 and 50 per cent.
In the limit of wide binaries, Eq.~\ref{eq:separate-rate} predicts increases of 11 and 36 per cent, respectively.
The observed increase in accretion above the equal-mass rate was most extreme at slow centre-of-mass velocities and is weakly positively correlated with binary separation.
The correlation between accretion rate and separation is unexpected, and partially counters the relations in \S\ref{sec:face-on-twins}, in which the accretion rates decreased at large binary separations.
This could perhaps be interpreted as the smaller star in a close binary disrupting the BHL flow, leading to a decrease in accretion rate.

In cases of unequal-mass binaries, the accretion rates of the individual stars are no longer expected to be equal.
Equation \ref{eq:separate-rate} also allows us to predict how the mass ratio of the binary evolves as each of the stars accretes material,
\begin{equation}
    \dot{q} \propto q \frac{q-1}{1+q}~.
    \label{eq:q-evolution}
\end{equation}
Since $\dot{q}$ is always negative, this implies that the accretion should be oligarchic, and that the initially more massive component will eventually dominate the system.
We can test this prediction by measuring the ratio $\dot{M}_2 / \dot{M}_1$ in our simulations; according to equation \ref{eq:separate-rate}, this should be equal to $q^{2}$.
Larger ratios correspond to more equal division of gas accreted by the binary, while lower values imply that the accretion is dominated by the larger component.
We find that in all of our simulations, $\dot{M}_2 / \dot{M}_1 > q^{2}$; i.e. that the measured accretion rates are more equal than predicted by the analytic theory.
However, the rate of accretion onto the less massive star is never sufficient to make $\dot{q}$ positive.

\subsection{Binary orbit}
\label{section:effect-on-orbit}
The evolution of the binary orbit depends on both the mass and angular momentum accreted by the binary. While so far our analysis has focussed on mass accretion, the angular momentum accretion rates can similarly be extracted from our simulations.
Because our simulations do not include dynamical friction, all changes in the orbital elements are due to accretion of mass and angular momentum by the stars.
Likewise, we can also measure the change in the orbital elements $a$ and $e$.
While the simulations are too short to determine long-term evolution, we can compute instantaneous rates of change.

\begin{figure*}
    \centering
    \includegraphics[width=\textwidth]{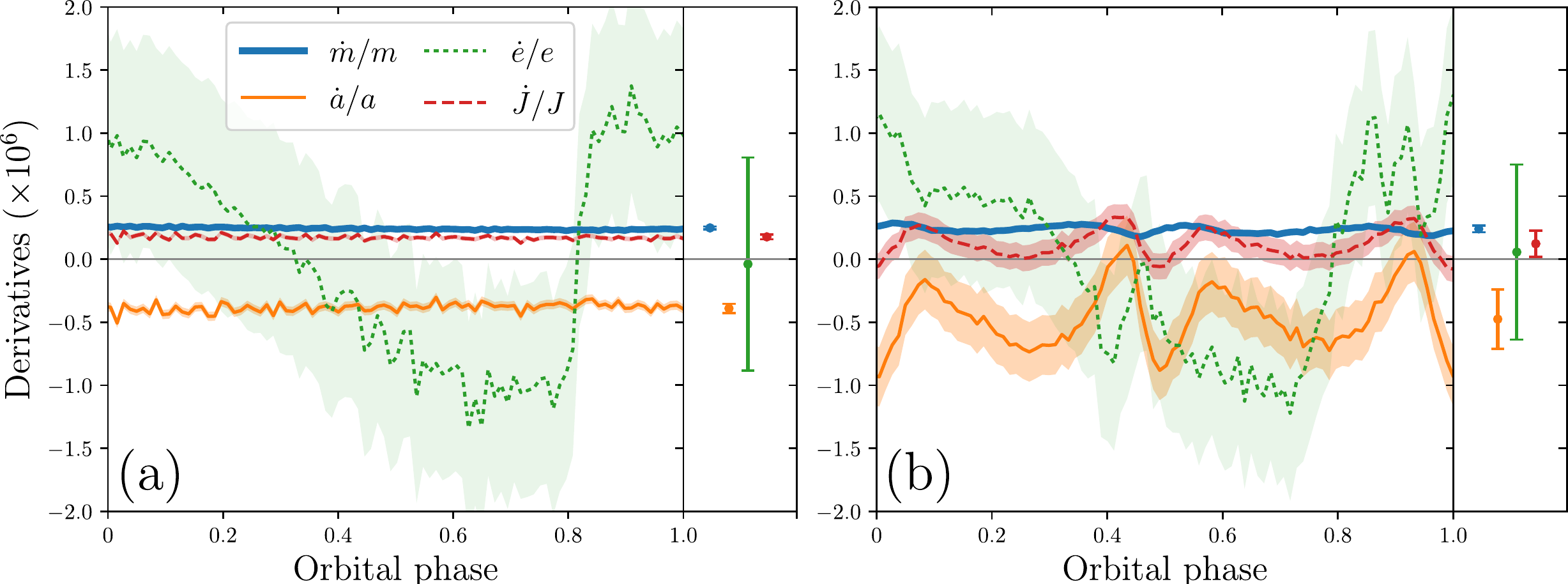}
    \caption{Derivatives of properties of the binary orbit (mass $m$, semi-major axis $a$, eccentricity $e$, and angular momentum $J$), plotted over one orbital period, for a face-on binary (left) and a side-on binary (right). The points and bars to the right of each plot show the orbital averages of each derivative, as well as its standard deviation.}
    \label{fig:orbital-derivatives}
\end{figure*}

The angular momentum ($J$) of a circular orbit is given by,
\begin{equation}
    J \propto \sqrt{M^3a}~,
\end{equation}
which can be manipulated to give the rate of change of orbital separation,
\begin{equation}
    \frac{\dot{a}}{a} = \frac{2\dot{J}}{J} - \frac{3\dot{M}}{M}.
\end{equation}
As a result, the binary orbit either expands or contracts depending on the angular momentum and mass accretion rates.

The quantity $d\ln{\dot{J}}/ d\ln{\dot{M}}$ is the ratio of specific angular momenta of the accreted material and the binary; when this ratio is less than 1.5, the orbit shrinks.
Excluding simulations with insufficient resolution, we find typical values of this ratio to be in the range -0.1 to 0.2, with a slight negative correlation with the binary centre-of-mass velocity.
Because this ratio is ratio is less than 1.5, the binary orbit shrinks in all our simulations in which the accretion is sufficiently resolved.
On timescales much longer than the orbital period, the decrease in binary separation enhances the effect of BBHLA (Figure~\ref{fig:face-on-twin-rates}).

All the binaries in our simulations started on circular orbits, however they gained eccentricity up to $\sim10^{-6}$ in the time before the accretion rates reached a steady state.
Once accretion is steady, we find that the eccentricity varies approximately sinusoidally with a period equal to the binary orbit.
Interestingly, this variation occurs in both face-on and side-on binaries, although the initial eccentricity increase is smaller in face-on binaries.
The time-averaged $\dot{e}$ in the steady state is zero, although the duration of our simulations is too short to fully determine the secular variation in eccentricity.
While we did not perform simulations of initially eccentric binaries, it is perhaps reasonable to expect that BBHLA should reduce the orbital eccentricity, by analogy with BHLA within a binary system
in which $\dot{e}/e$ is generally negative \citep{edot1,edot2}.

\section{Discussion}
In this section, we discuss the implications and applications of this work, and compare to existing literature.

\subsection{Comparison to existing work}

The primary pre-existing study of BBHLA is \cite{Soker04} (hereafter S04), in which the author determines that when a binary orbits inside the wind from a tertiary companion, the accreted material forms a disk with sufficient angular momentum to launch jets.
It is difficult to draw direct comparisons between that work, which focussed on angular momentum accretion onto a binary orbiting a companion, and our work, which concerns mass accretion and orbital evolution in idealised circumstances.
We can, however test the assumptions to determine the extent to which the simple analytic model is realistic.

The analysis in S04 is predicated on the assumed ordering of length scales,
\begin{equation}
    R_1 \ll w \lesssim a \ll r_{\mathrm{a}} \ll a_{\mathrm{trip}}~,
\end{equation}
where $R_1$ is the radius of one of the stars in the binary, $w$ is the width of the Bondi-Hoyle accretion column, $a$ is the binary separation ($a_{12}$ in S04), $r_{\mathrm{a}}$ is the BHLA radius ($R_{\mathrm{acc}}$ in S04), and $a_{\mathrm{trip}}$ the distance to the tertiary companion (which may be treated as infinite in our analysis).
Our simulations verify that when $a \ll r_{\mathrm{a}}$, the binary produces a single combined bow shock, and that the form of the flow is similar to the single star case outside the region within a distance about $a$ of the binary.

However, our simulations do not display the narrow column described in \cite{HoyleLyttleton}; the gas approaches the binary from a wider region.
This explains our resulting eccentricity variation with orbital phase. During half the orbit the star encounters material moving in the opposite direction to its orbital motion, slowing the star and increasing eccentricity, while during the other half of the orbit, the accreted material moves in the same direction as the star, reducing the eccentricity.
If, instead, the material arrives in a narrow stream, we would observe strongly varying accretion rates and rapidly increasing eccentricity.

Our work is similar to \cite{LinMurray2007} who examine the accretion of gas onto stellar clusters encountering a slab of gas.
They find that the main effect depends on the ratio of cluster centre-of-mass velocity to the velocity dispersion within the cluster.
When the centre-of-mass velocity exceeds the velocity dispersion, the stars accrete individually, and the cluster as a whole retains little gas.
When the dispersion is higher, gas is accreted onto the cluster, but the stars move too quickly to accrete much individually. The gas is retained in the cluster, and accreted over time by the stars.
They consider accretion during a single encounter with a slab of gas; it is less clear what the effect of continuous accretion would be.

The ratio of cluster velocity to velocity dispersion is analogous to the ratio of binary centre-of-mass velocity and orbital velocity in our work.
The first case, where the centre-of-mass velocity is high, corresponds to our wide binaries, where the stars form separate bow shocks.
The second case, where the cluster is found to retain gas, does not appear to have an equivalent in our simulations; we have no cases where gas is retained by the binary without being quickly accreted by either star.
There are several possible reasons why we may not have seen this: the first is that many point masses may be required to retain gas, while we only have two.
Another possibility is that our simulations lack sufficient resolution to produce circumbinary or circumstellar disks.
In this case, formation of a circumbinary disk would be equivalent to the cluster retaining gas after the encounter with the slab, without gas becoming associated with the individual cluster stars.

A pair of recent papers, \cite{kaaz19} and \cite{antoni19}, consider BHLA onto clusters and binary stars, respectively.
They reach the same broad conclusions as us, namely that in systems with $a \gg r_{\mathrm{a}}$ the stars accrete separately, while at closer separations the system accretes gas more like a single body.
While we have used SPH and MFV, they use an adaptive mesh refinement (AMR) fluid simulation, so the fact that we reach similar conclusions implies that our results are largely free of spurious simulation artefacts.
The simulations in \cite{antoni19} are both performed at a higher resolution and run for a longer time than our simulations, allowing the authors to investigate the long term evolution of the binary orbit.
In all binaries that they simulate, they find that $\dot{v}/v < \dot{a}/a < \dot{m}/m$; that the binary should come to a halt relative to the gas before the orbit substantially shrinks or the total mass substantially increases.
Our simulations agree with the ordering and approximate ratio of $\dot{a}/a$ and $\dot{m}/m$, however, we do not calculate $\dot{v}/v$, due to the shorter duration of our simulations, and lack of gas drag on the stars.
In the astrophysical contexts to which we plan to apply these results, there are additional sources of momentum which can re-accelerate the binary after its centre-of-mass velocity has decreased.
In a cluster, this acceleration could be due to a multi-body interaction which does not disrupt the binary; in a triple star system, the lost momentum can be regained from the orbit of the binary about the tertiary companion.

\subsection{Applications}
We have shown that by being associated in a binary, two stars may accrete up to twice as much material as if they were isolated.
This has implications for several astrophysical situations.
In young globular clusters, stars may accrete gas left over from star formation, or ejected by intermediate-mass stars during their AGB phase.
This gas is a possible source of $\alpha$-rich elements observed in the spectra of ``second generation'' stars \citep{bastian18}.
Since binary stellar systems are, on average, more massive than single stars, mass segregation results in binaries accumulating towards the centre of the cluster, in close proximity to the massive AGB stars and residual gas.
In these regions, the rate of three-body encounters is highest, so it is unlikely that the binary system remains bound for long \citep{ivanova05}.
Whether the woulds last long enough to substantially affect the accretion of material onto its constituent stars is dependent on the properties of the cluster.

A similar case can be made for stars near the Galactic centre, with similar caveats regarding the rate of multiple-body encounters which could unbind the binary.
BHLA onto binaries orbiting within a gas disk affects its orbit around the Galactic centre \citep{Baruteau11}, and accretion onto black hole binaries may substantially alter their mass and prompt a gravitational-wave-producing merger \citep{Yi18}.

The final case in which we consider the role of binary BHLA is in wind mass transfer in hierarchical triple stars, as described by \cite{Soker04}.
In this case, a companion to a binary produces a wind which encounters the binary.
The assumptions we make in this paper are most valid in the case where the system is very hierarchical, i.e. that the binary orbital separation is much less than the distance to the third star.
We also assume that incoming gas is uniform, with no velocity or density gradients.
This assumption not only requires the large separation mentioned above, but also that the wind is supersonic and only weakly focussed through the central Lagrangian point.

\subsection{Limitations and future work}
While our simulations are sufficient to determine general trends, this study has been primarily limited by the time and computing power requirements.
We elected to study a wide region of parameter space with relatively low-resolution simulations, meaning that there are cases where the phenomena we were looking for became unresolved, and that the simulations could not run for long enough to determine the long-term evolution.
For systems which appeared to be only marginally resolved, an increase in simulation resolution by around a factor of two would be sufficient to fully resolve the bow shock, although this increases the computational complexity by a factor of at least eight.
One important phenomenon we have neglected is the formation of disks, both around the binary and around the individual stars.
Our simulations lacked the resolution to accurately form accretion disks, although their formation is expected due to the angular momentum of the infalling material.
As a result, we have been unable to draw definitive conclusions regarding the angular momentum accretion onto the individual stars.

We have also not accounted for radiative cooling, molecular dissociation or ionisation. These processes are sinks of thermal energy and alter the properties of the accreting material, but complicate the modelling due to their dependence on the gas density and properties of the star.

Similarly, a more complete treatment of this problem could include the effects of a magnetic field. Unsurprisingly, the addition of magnetic fields has the potential to complicate the simulations, whether the field originates in the interstellar medium, from one or both of the accreting stars, or the donor star in a triple system.
In \cite{lee14}, the authors perform simulations of BHLA of molecular gas with a magnetic field, finding that in the case of Bondi accretion, the accretion rate is significantly reduced when the magnetic pressure exceeds one per cent of the gas pressure.
For moving accretors, the critical magnetic pressure increases roughly proportionally to the accretor's Mach number, so magnetic fields will have less of an effect on fast accretors.

Applying the criteria in \cite{lee14}, the condition on the ambient magnetic field to not disrupt the accretion is,

\begin{equation}
    B \lesssim 3.23 \times 10^{-5}\,\mathrm{T} \left( \frac{T}{100\,\mathrm{K}}\right)^{1/2} \left( \frac{\rho}{10^{-14}~\mathrm{kg\,m}^{-3}}\right)^{1/2} (1 + \mathcal{M}^2)^{1/2},
    \label{eq:magnetic-field-limit}
\end{equation}
where $\mathcal{M}$ is the accretor's Mach number.

To determine whether BHLA can be magnetically inhibited in a multiple stellar system, we can examine the example case of a companion accreting from the wind of a giant star with a radius of $R = 10\,R_{\odot}$ and a magnetic field with a surface strength of $0.01\,\mathrm{T}$ corresponding to the strongest magnetic fields measured in G/K giants by \citet{2015A&A...574A..90A}.

The field strength a distance $r$ from the giant is, assuming a dipole field geometry, approximately,

\begin{equation}
    B = 0.01\,\mathrm{T}~ \left( \frac{10\,R_\odot}{r}
\right)^3 = 10\,\mathrm{T}~ \frac{R^{3}_\odot}{r^3}\,,
\end{equation}
hence at a minimum distance to the accreting star of $r=2R=20\,R_\odot$ then $B=1/8\times 10^{-2}\,\mathrm{T}=1.25\times 10^{-3}\,\mathrm{T}$.
Adopting $T = 1000\,\mathrm{K}$ and $\rho = 10^{-10}\,\mathrm{kg\,m}^{-3}$ in the vicinity of the accretor, Equation~\ref{eq:magnetic-field-limit} gives a threshold magnetic field strength of at least $0.01\,(1+\mathcal{M}^2)\,\mathrm{T}$.
As a result, we would not expect BHLA to be strongly affected by the presence of the magnetic field in a typical red-giant wind accretion scenario, even for low Mach numbers.

In realistic physical scenarios, the accreted material will have some angular momentum, so formation of a disk around either or both of the accretors.
In these cases, the interaction of the disk with magnetic fields can reduce the accretion rate and produce outflows \citep{moscibrodzka09}.

There are many directions in which one could expand on this work. In addition to wider and finer sampling of the parameter space, and higher resolution simulations, one could also introduce more varied physical context.
One possible route would be to abandon the assumption of a homogeneous medium, and allow the incoming material to have variations in density and temperature (in directions both parallel and perpendicular to the relative velocity;  \citealt{macleod14}).
Introducing a longitudinal density gradient and angular momentum would produce a model that better represents accretion in a triple stellar system, while a transverse gradient would represent the model in \cite{Soker16}, in which an asymptotic giant branch star swallows a binary that merges inside the envelope of the larger star.

\section{Conclusions}
Using the code \textsc{gandalf}, we have performed a suite of simulations of Bondi-Hoyle-Lyttleton accretion (BHLA) onto single and binary accretors, with a range of separations, velocities, inclinations and mass ratios, using SPH and MFV schemes.
We recover the analytic form of the BHLA rate in single stars, and find the expected result that in very close binaries the accretion rate approaches the single-star rate.
In wide binaries the mass accretion rate drops to half the single-star rate, where in slow-moving binaries, the rate reaches the asymptote at lower binary separations.

We find that the inclination of the binary relative to the centre-of-mass velocity vector has comparatively little effect on the mass accretion rates but is important when considering the orbital evolution of the binary because it alters the angular momentum accretion rate.
In binaries where the stars have unequal masses, the accretion rate of the less massive star is enhanced by its proximity to the companion, and the more massive star accretes at a disproportionately higher rate, leading to a decrease in the mass ratio, $q$.
If this trend continues over timescales in which the stars' masses increase substantially, it implies that binaries should evolve away from equal mass ratios toward systems that are dominated by a single component.

While this initial study was limited by available resolution and computation time, we have demonstrated the general trends in accretion rates as a function of binary separation, velocity, inclination, and mass ratio.
Simulations with higher spatial resolution would demonstrate the behaviour of the gas in the immediate vicinity of the stars, including the potential formation of circumstellar and circumbinary discs. Simulating many binary orbits would also constrain the long-term orbital evolution. 
Future work could involve determining the role of eccentricity in binary BHLA, as well as more physically realistic simulations of the applications in globular clusters, dense stellar environments and multiple stellar systems.

\section*{Acknowledgements}
We thank the referee for their constructive comments which improved this work, and Jonathan Mackey, Zhengwei Liu and Noam Soker for useful discussions and inspiration. This work was performed using the Cambridge Service for Data Driven Discovery (CSD3), part of which is operated by the University of Cambridge Research Computing, and the DiRAC Data Centric system at Durham University, operated by the Institute for Computational Cosmology, both on behalf of the STFC DiRAC HPC Facility (www.dirac.ac.uk). The Data Centric System was funded by a BIS National E-infrastructure capital grant ST/K00042X/1, STFC capital grant ST/K00087X/1, DiRAC Operations grant ST/K003267/1 and Durham University. DiRAC is part of the National e-Infrastructure. RGI thanks the STFC for Rutherford fellowship funding under grant number ST/L003910/1. Hello to Jason Isaacs. G.R. acknowledges support from the Netherlands Organisation for Scientific Research (NWO, program number 016.Veni.192.233).

Figures were produced using Matplotlib \citep{matplotlib}, SPLASH \citep{SPLASH}, and Inkscape (\url{https://inkscape.org}).




\bibliographystyle{mnras}

\bibliography{refs}



\clearpage 
\appendix

\section{Resolution study}
\label{app:resolution-study}
To ensure that our simulations are adequately resolved, we performed a resolution study consisting of a series of simulations with varying resolution and accretor velocity, with the aim of comparing our results with those of \citet[hereafter R94]{Ruffert94}.

In R94 and its subsequent papers, the authors derive an expression for the BHL accretion rate (their Eq.~15) which interpolates between cases when the flow is well- and poorly-resolved.
They then go on to test their analytic rates using a set of adaptive mesh refinement (AMR) simulations, and find them to be in good agreement.

Qualitatively, they find that when the accretion is well-resolved, the bow shock is separated from the accretor, and there is little to no resolution-dependence in the accretion rates.
However for larger accretors, the shock cone becomes attached to the sink particle, yielding unreliable accretion rates that increase proportionally with the square of the accretor radius.
The transition between these regimes depends on the velocity of the accretor velocity, with faster-moving accretors becoming unresolved at smaller physical scales.

If the accretor size is held at about $0.016$ Bondi radii, the default in our main simulations, the results of R94 imply that accretion is unresolved at velocities of 6 or higher, which agrees well with our results (Fig.~\ref{fig:single-star-rates}).

\begin{figure}
    \centering
    \includegraphics[width=\columnwidth]{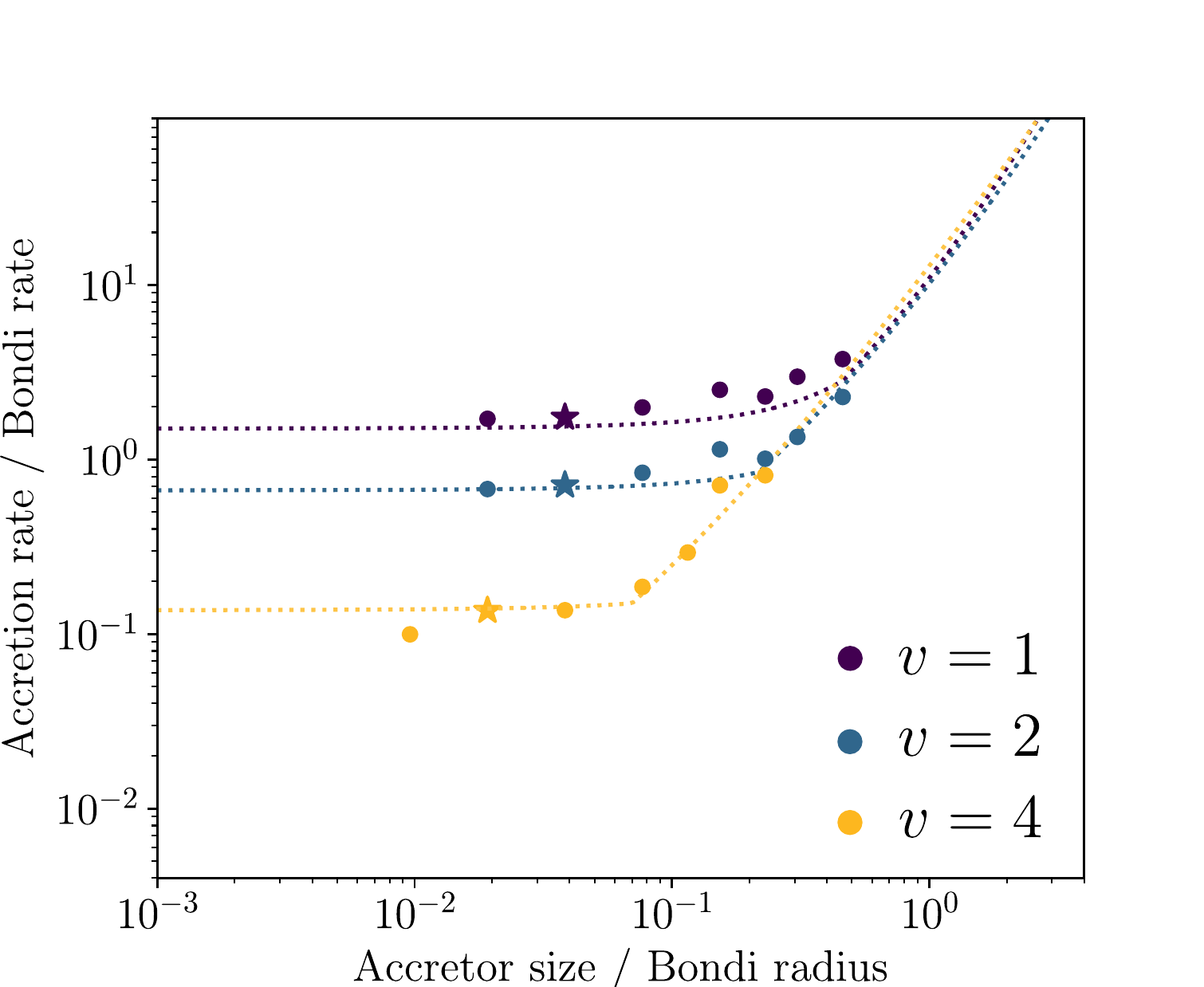}
    \caption{Our version of Fig.~6 in \citet{Ruffert94}, showing the dependence of the accretion rates on the physical size of the accretor (normalised to the Bondi rate and radius respectively). The three colours correspond to different accretor velocities, and the dotted lines are Eq.~15 from R94. Points marked with stars indicate the standard resolution used in our single-star simulations.}
    \label{fig:resolution-study}
\end{figure}

The simulations in this resolution study span a range of accretor velocities between Mach 1 and 4, and accretor sizes between $0.01$ and $0.5$ stationary Bondi radii.
At higher resolutions than this, the computational cost increases rapidly, while at lower resolutions, our simulation fail to converge.
The results of our simulations match those of R94 well (Fig.~\ref{fig:resolution-study}); at the three different accretor velocities we observe both the flat, well-resolved region and the slope up to the unresolved rate for large accretors, with the transition point at the size predicted in R94.

Our higher-resolution accretion rates generally fall below the analytic predictions by between 10 and 20 percent, which is consistent with other simulations of BHLA \citep{Edgar04}.

\begin{figure*}
    \centering
    \includegraphics[width=\textwidth]{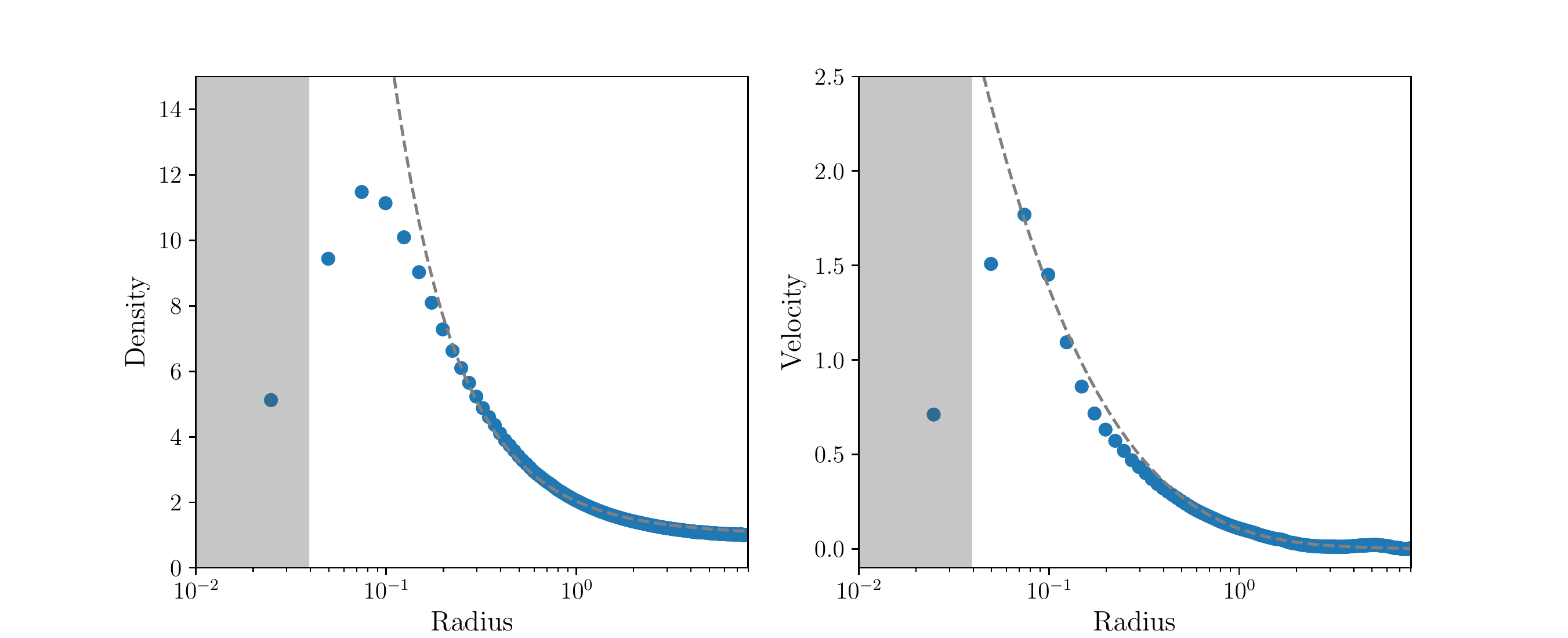}
    \caption{Density and velocity profiles of spherical Bondi accretion for simulation 0009. The grey region to the left represents the radius of the sink particle; analytic profiles from \citet{Bondi} are also plotted in grey. Density and velocity are normalised to the ambient density and sound speed; the radius is in simulation units, in which the sink particle has a radius of 0.039 and the Bondi radius (\ref{eq:bondi-radius}) is 0.6.}
    \label{fig:bondi-profiles}
\end{figure*}

We also compare the results of our simulation 0009 to the expected density and velocity profiles of spherical Bondi accretion (Figure~\ref{fig:bondi-profiles}).
The analytic treatment in \cite{Bondi} shows that for an adiabatic index of $5/3$, the fluid density and velocity increase without bound near the accretor, but the flow never becomes supersonic.
Our simulations do show a sonic surface within one to two smoothing lengths of the sink, which deviates from the theory, but does ensure that the conditions immediately adjacent do not affect the extended flow. The cause of the extra acceleration is the artificial pressure gradient created when particles are removed from the simulation during accretion.

\section{List of simulation parameters}
\onecolumn
{

\begin{longtable}{cccccccccccc}
    \caption{List of our SPH simulation runs. `Resolution' refers to the size of the simulation domain in units of the SPH particle spacing while `dimension' is the absolute size in simulation units. $x$, $y$, and $z$ specify the quantity along the corresponding coordinate axes; the $y$ and $z$ values are equal in all cases. The next four columns list the binary mass ratio ($q$), inclination ($i$), semi-major axis ($a$), and centre-of-mass velocity ($v$). Generally, the four digits of the simulation ID correspond to the values of $q$, $i$, $a$, and $v$, respectively. There are some exceptions to this, as more than 10 different velocities were used. The final three columns contain the total binary accretion rate, normalised by the ambient gas density ($\dot{M}/\rho$); ratio of accretion rates of each of the stars ($\dot{M}_1/\dot{M}_2$, where $M_1$ is the more massive star); and the fractional change of semi-major axis, normalised by gas density ($\dot{a}/(\rho a)$). Simulations marked with a dagger ($^{\dagger}$) were repeated using an MFV simulation.}
    \label{tab:parameters-table}
    \\\hline
    \multirow{2}{*}{ID} & \multicolumn{2}{c}{Resolution} & \multicolumn{2}{c}{Dimension} & \multirow{2}{*}{$1/q$} & \multirow{2}{*}{$i(^{\circ})$} & \multirow{2}{*}{$a$} & \multirow{2}{*}{$v$} & \multirow{2}{*}{$\dot{M}/\rho$} & \multirow{2}{*}{$\dot{M}_1/\dot{M}_2$} & \multirow{2}{*}{$\dot{a}/(\rho a)$}\\
    & $x$ & $y,z$ & $x$ & $y,z$ & & & & & & & \\\hline
0000 & 512 & 128 & 16 & 4 & - & - & - & 1 & 0.880 &  - &  - \\
0001$^{\dagger}$ & 512 & 128 & 16 & 4 & - & - & - & 2 & 0.356 &  - &  - \\
0002 & 512 & 128 & 16 & 4 & - & - & - & 4 & 0.0683 &  - &  - \\
0004 & 256 & 128 & 8 & 4 & - & - & - & 0.5 & 0.979 &  - &  - \\
0005 & 512 & 128 & 16 & 4 & - & - & - & 3 & 0.118 &  - &  - \\
0006 & 512 & 128 & 16 & 4 & - & - & - & 5 & 0.0389 &  - &  - \\
0007 & 512 & 96 & 16 & 3 & - & - & - & 7 & 0.0354 &  - &  - \\
0008 & 512 & 96 & 16 & 3 & - & - & - & 6 & 0.0316 &  - &  - \\
0009 & 128 & 128 & 4 & 4 & - & - & - & 0 & 0.919 &  - &  - \\
0009b & 203 & 203 & 8 & 8 & - & - & - & 0 & 1.008 &  - &  - \\
0009c & 322 & 322 & 16 & 16 & -& - & - & 0 & 1.090 & - & - \\
0010 & 256 & 128 & 8 & 4 & 1 & 90 & 1 & 1 & 0.569 & 0.999 & -1.14\\
0011 & 512 & 128 & 16 & 4 & 1 & 90 & 1 & 2 & 0.167 & 1.000 & -0.31\\
0012 & 512 & 128 & 16 & 4 & 1 & 90 & 1 & 4 & 0.0666 & 0.997 & -0.22\\
0014 & 512 & 128 & 16 & 4 & 1 & 90 & 1 & 2.5 & 0.108 & 1.000 & -0.3\\
0015 & 512 & 128 & 16 & 4 & 1 & 90 & 1 & 1.5 & 0.369 & 1.000 & -0.52\\
0019 & 128 & 128 & 4 & 4 & 1 & - & 1 & 0 & 0.546 & 1.000 & -0.97\\
0020 & 256 & 128 & 8 & 4 & 1 & 90 & 1/2 & 1 & 0.616 & 1.000 & -2.55\\
0021 & 512 & 128 & 16 & 4 & 1 & 90 & 1/2 & 2 & 0.248 & 1.000 & -0.78\\
0022 & 512 & 128 & 16 & 4 & 1 & 90 & 1/2 & 4 & 0.0654 & 1.000 & -0.45\\
0029 & 128 & 128 & 4 & 4 & 1 & - & 1/2 & 0 & 0.580 & 1.000 & -2.19\\
0030 & 256 & 128 & 8 & 4 & 1 & 90 & 1/4 & 1 & 0.685 & 1.000 & -6.57\\
0031 & 512 & 128 & 16 & 4 & 1 & 90 & 1/4 & 2 & 0.335 & 1.000 & -3.77\\
0032 & 512 & 128 & 16 & 4 & 1 & 90 & 1/4 & 4 & 0.0674 & 1.010 & -0.92\\
0039 & 128 & 128 & 4 & 4 & 1 & - & 1/4 & 0 & 0.629 & 1.000 & -5.51\\
0040 & 256 & 128 & 8 & 4 & 1 & 90 & 1/8 & 1 & 0.786 & 1.000 & -17.81\\
0041 & 512 & 128 & 16 & 4 & 1 & 90 & 1/8 & 2 & 0.359 & 1.000 & -8.47\\
0042 & 512 & 128 & 16 & 4 & 1 & 90 & 1/8 & 4 & 0.0729 & 1.000 & -1.92\\
0049 & 128 & 128 & 4 & 4 & 1 & - & 1/8 & 0 & 0.759 & 1.000 & -16.29\\
0050 & 256 & 128 & 8 & 4 & 1 & 90 & 1/16 & 1 & 0.953 & 1.000 & -49.14\\
0051 & 512 & 128 & 16 & 4 & 1 & 90 & 1/16 & 2 & 0.401 & 1.000 & -22.68\\
0052 & 512 & 128 & 16 & 4 & 1 & 90 & 1/16 & 4 & 0.082 & 1.000 & 7.96\\
0059 & 128 & 128 & 4 & 4 & 1 & - & 1/16 & 0 & 0.987 & 1.000 & -50.28\\
0070 & 256 & 128 & 8 & 4 & 1 & 90 & 0.7 & 1 & 0.604 & 1.000 & -1.85\\
0071 & 512 & 128 & 16 & 4 & 1 & 90 & 0.7 & 2 & 0.194 & 0.999 & -0.47\\
0072 & 512 & 128 & 16 & 4 & 1 & 90 & 0.7 & 4 & 0.068 & 1.000 & -0.33\\
0079 & 128 & 128 & 4 & 4 & 1 & - & 0.7 & 0 & 0.567 & 1.000 & -1.44\\
0110 & 256 & 128 & 8 & 4 & 1 & 0 & 1 & 1 & 0.532 & 0.911 & -0.79\\
0111 & 512 & 128 & 16 & 4 & 1 & 0 & 1 & 2 & 0.230 & 0.681 & -0.51\\
0112 & 512 & 128 & 16 & 4 & 1 & 0 & 1 & 4 & 0.0631 & 1.000 & -0.23\\
0114 & 512 & 128 & 16 & 4 & 1 & 0 & 1 & 2.5 & 0.135 & 0.718 & -0.39\\
0115 & 512 & 128 & 16 & 4 & 1 & 0 & 1 & 1.5 & 0.375 & 0.967 & -0.45\\
0120 & 512 & 128 & 16 & 4 & 1 & 0 & 1/2 & 1 & 0.566 & 0.963 & -2.56\\
0121$^{\dagger}$ & 512 & 128 & 16 & 4 & 1 & 0 & 1/2 & 2 & 0.240 & 0.944 & -0.95\\
0122 & 512 & 128 & 16 & 4 & 1 & 0 & 1/2 & 4 & 0.0691 & 1.030 & -0.46\\
0130 & 256 & 128 & 8 & 4 & 1 & 0 & 1/4 & 1 & 0.651 & 0.997 & -5.85\\
0131 & 512 & 128 & 16 & 4 & 1 & 0 & 1/4 & 2 & 0.283 & 0.994 & -2.42\\
0132 & 512 & 128 & 16 & 4 & 1 & 0 & 1/4 & 4 & 0.0756 & 1.010 & -1.09\\
0140 & 256 & 128 & 8 & 4 & 1 & 0 & 1/8 & 1 & 0.803 & 1.000 & -17.73\\
0141 & 512 & 128 & 16 & 4 & 1 & 0 & 1/8 & 2 & 0.344 & 0.997 & -8.4\\
0142 & 512 & 128 & 16 & 4 & 1 & 0 & 1/8 & 4 & 0.0855 & 0.976 & -2.47\\
0170 & 256 & 128 & 8 & 4 & 1 & 0 & 0.7 & 1 & 0.546 & 1.010 & -1.24\\
0171 & 512 & 128 & 16 & 4 & 1 & 0 & 0.7 & 2 & 0.235 & 0.914 & -0.72\\
0172 & 512 & 128 & 16 & 4 & 1 & 0 & 0.7 & 4 & 0.0665 & 0.868 & -0.3\\
0210 & 256 & 128 & 8 & 4 & 1 & 45 & 1 & 1 & 0.553 & 0.939 & -0.99\\
0211 & 512 & 128 & 16 & 4 & 1 & 45 & 1 & 2 & 0.207 & 0.730 & -0.39\\
0212 & 512 & 128 & 16 & 4 & 1 & 45 & 1 & 4 & 0.0673 & 0.964 & -0.24\\
0220 & 256 & 128 & 8 & 4 & 1 & 45 & 1/2 & 1 & 0.577 & 0.944 & -2.54\\
0221 & 512 & 128 & 16 & 4 & 1 & 45 & 1/2 & 2 & 0.253 & 0.908 & -1.0\\
0222 & 512 & 128 & 16 & 4 & 1 & 45 & 1/2 & 4 & 0.0673 & 0.997 & -0.46\\
0230 & 256 & 128 & 8 & 4 & 1 & 45 & 1/4 & 1 & 0.667 & 1.000 & -6.37\\
0231 & 512 & 128 & 16 & 4 & 1 & 45 & 1/4 & 2 & 0.304 & 1.030 & -3.02\\
0232 & 512 & 128 & 16 & 4 & 1 & 45 & 1/4 & 4 & 0.0687 & 0.997 & -0.95\\
0240 & 256 & 128 & 8 & 4 & 1 & 45 & 1/8 & 1 & 0.806 & 0.999 & -18.05\\
0241 & 512 & 128 & 16 & 4 & 1 & 45 & 1/8 & 2 & 0.350 & 0.999 & -8.57\\
0242 & 512 & 128 & 16 & 4 & 1 & 45 & 1/8 & 4 & 0.0786 & 0.990 & -2.08\\
0270 & 256 & 128 & 8 & 4 & 1 & 45 & 0.7 & 1 & 0.567 & 0.990 & -1.6\\
0271 & 512 & 128 & 16 & 4 & 1 & 45 & 0.7 & 2 & 0.222 & 0.928 & -0.59\\
0272 & 512 & 128 & 16 & 4 & 1 & 45 & 0.7 & 4 & 0.0666 & 0.974 & -0.32\\
1010 & 256 & 128 & 8 & 4 & 2 & 90 & 1 & 1 & 0.553 & 2.770 & -0.9\\
1011 & 512 & 128 & 16 & 4 & 2 & 90 & 1 & 2 & 0.198 & 3.670 & -0.36\\
1012 & 512 & 128 & 16 & 4 & 2 & 90 & 1 & 4 & 0.070 & 1.600 & -0.25\\
1019 & 128 & 128 & 4 & 4 & 2 & - & 1 & 0 & 0.607 & 2.940 & -1.12\\
1020 & 256 & 128 & 8 & 4 & 2 & 90 & 1/2 & 1 & 0.651 & 2.500 & -2.75\\
1021 & 512 & 128 & 16 & 4 & 2 & 90 & 1/2 & 2 & 0.264 & 3.130 & -0.88\\
1022 & 512 & 128 & 16 & 4 & 2 & 90 & 1/2 & 4 & 0.0693 & 1.610 & -0.5\\
1029 & 128 & 128 & 4 & 4 & 2 & - & 1/2 & 0 & 0.677 & 2.520 & -2.79\\
1030 & 256 & 128 & 8 & 4 & 2 & 90 & 1/4 & 1 & 0.706 & 2.180 & -6.79\\
1031 & 512 & 128 & 16 & 4 & 2 & 90 & 1/4 & 2 & 0.334 & 1.990 & -3.68\\
1032 & 512 & 128 & 16 & 4 & 2 & 90 & 1/4 & 4 & 0.0683 & 1.500 & -0.98\\
1039 & 128 & 128 & 4 & 4 & 2 & - & 1/4 & 0 & 0.684 & 2.320 & -6.09\\
1040 & 256 & 128 & 8 & 4 & 2 & 90 & 1/8 & 1 & 0.799 & 1.800 & -18.62\\
1041 & 512 & 128 & 16 & 4 & 2 & 90 & 1/8 & 2 & 0.359 & 1.580 & -8.83\\
1042 & 512 & 128 & 16 & 4 & 2 & 90 & 1/8 & 4 & 0.0736 & 1.470 & -2.08\\
1049 & 128 & 128 & 4 & 4 & 2 & - & 1/8 & 0 & 0.810 & 1.870 & -18.35\\
2010 & 256 & 128 & 8 & 4 & 4 & 90 & 1 & 1 & 0.687 & 6.880 & -1.29\\
2011 & 512 & 128 & 16 & 4 & 4 & 90 & 1 & 2 & 0.249 & 9.950 & -0.44\\
2012 & 512 & 128 & 16 & 4 & 4 & 90 & 1 & 4 & 0.0737 & 2.280 & -0.31\\
2019 & 128 & 128 & 4 & 4 & 4 & - & 1 & 0 & 0.704 & 8.050 & -1.29\\
2020 & 256 & 128 & 8 & 4 & 4 & 90 & 1/2 & 1 & 0.720 & 5.740 & -3.08\\
2021 & 512 & 128 & 16 & 4 & 4 & 90 & 1/2 & 2 & 0.282 & 9.750 & -0.97\\
2022 & 512 & 128 & 16 & 4 & 4 & 90 & 1/2 & 4 & 0.0742 & 2.180 & -0.64\\
2029 & 128 & 128 & 4 & 4 & 4 & - & 1/2 & 0 & 0.723 & 6.520 & -2.78\\
2030 & 256 & 128 & 8 & 4 & 4 & 90 & 1/4 & 1 & 0.767 & 4.420 & -7.78\\
2031 & 512 & 128 & 16 & 4 & 4 & 90 & 1/4 & 2 & 0.344 & 3.870 & -3.82\\
2032 & 512 & 128 & 16 & 4 & 4 & 90 & 1/4 & 4 & 0.0735 & 2.100 & -1.26\\
2039 & 128 & 128 & 4 & 4 & 4 & - & 1/4 & 0 & 0.757 & 4.980 & -7.22\\
2110 & 256 & 128 & 8 & 4 & 4 & 0 & 1 & 1 & 0.643 & 6.950 & -1.13\\
2111 & 512 & 128 & 16 & 4 & 4 & 0 & 1 & 2 & 0.267 & 4.980 & -0.4\\
2112 & 512 & 128 & 16 & 4 & 4 & 0 & 1 & 4 & 0.0711 & 2.540 & 0.1\\
2120 & 256 & 128 & 8 & 4 & 4 & 0 & 1/2 & 1 & 0.675 & 6.160 & -2.62\\
2121$^{\dagger}$ & 512 & 128 & 16 & 4 & 4 & 0 & 1/2 & 2 & 0.300 & 5.920 & -1.13\\
2122 & 512 & 128 & 16 & 4 & 4 & 0 & 1/2 & 4 & 0.0759 & 2.140 & -0.72\\
2130 & 256 & 128 & 8 & 4 & 4 & 0 & 1/4 & 1 & 0.746 & 4.530 & -7.08\\
2131 & 512 & 128 & 16 & 4 & 4 & 0 & 1/4 & 2 & 0.328 & 4.690 & -3.12\\
2132 & 512 & 128 & 16 & 4 & 4 & 0 & 1/4 & 4 & 0.0792 & 2.050 & -1.6\\
2210 & 256 & 128 & 8 & 4 & 4 & 45 & 1 & 1 & 0.664 & 6.730 & -1.22\\
2211 & 512 & 128 & 16 & 4 & 4 & 45 & 1 & 2 & 0.263 & 6.060 & -0.38\\
2220 & 256 & 128 & 8 & 4 & 4 & 45 & 1/2 & 1 & 0.704 & 5.990 & -2.9\\
2221 & 512 & 128 & 16 & 4 & 4 & 45 & 1/2 & 2 & 0.297 & 5.550 & -1.27\\
2222 & 512 & 128 & 16 & 4 & 4 & 45 & 1/2 & 4 & 0.0744 & 2.210 & -0.64\\
2230 & 256 & 128 & 8 & 4 & 4 & 45 & 1/4 & 1 & 0.746 & 4.360 & -7.42\\
2231 & 512 & 128 & 16 & 4 & 4 & 45 & 1/4 & 2 & 0.334 & 4.470 & -3.61\\
2232 & 512 & 128 & 16 & 4 & 4 & 45 & 1/4 & 4 & 0.0756 & 2.140 & -1.32\\
    \hline
    \end{longtable}
}

\twocolumn


\bsp	
\label{lastpage}
\end{document}